\documentclass[preprint,notoc]{JHEP3}
\usepackage{graphicx}
\usepackage{amsmath}
\usepackage{psfrag}
\usepackage[sort&compress,numbers]{natbib}

\def\be{\begin{equation}}
\def\ee{\end{equation}}
\def\bea{\begin{eqnarray}}
\def\eea{\end{eqnarray}}

\def\eqn#1{eq.~(\ref{#1})}

\def\e{\epsilon}

\def\la{\langle}

\def\Psl{\not{\hbox{\kern-2.3pt $P$}}}
\def\psl{\not{\hbox{\kern-2.3pt $p$}}}
\def\qsl{\not{\hbox{\kern-2.3pt $q$}}}
\def\Ksl{\not{\hbox{\kern-2.3pt $K$}}}
\def\ksl{\not{\hbox{\kern-2.3pt $k$}}}
\def\esl{\not{\hbox{\kern-2.3pt $\pol$}}}

\def\pol{\varepsilon}

\def\spa#1.#2{\left\langle#1\,#2\right\rangle}
\def\spb#1.#2{\left[#1\,#2\right]}

\def\lor#1.#2{\left(#1\,#2\right)}
\def\sand#1.#2.#3{%
\left\langle\smash{#1}{\vphantom1}^{-}\right|{#2}%
\left|\smash{#3}{\vphantom1}^{-}\right\rangle}
\def\sandp#1.#2.#3{%
\left\langle\smash{#1}{\vphantom1}^{-}\right|{#2}%
\left|\smash{#3}{\vphantom1}^{+}\right\rangle}
\def\sandpp#1.#2.#3{%
\left\langle\smash{#1}{\vphantom1}^{+}\right|{#2}%
\left|\smash{#3}{\vphantom1}^{+}\right\rangle}
\def\sandpm#1.#2.#3{%
\left\langle\smash{#1}{\vphantom1}^{+}\right|{#2}%
\left|\smash{#3}{\vphantom1}^{-}\right\rangle}
\def\sandmp#1.#2.#3{%
\left\langle\smash{#1}{\vphantom1}^{-}\right|{#2}%
\left|\smash{#3}{\vphantom1}^{+}\right\rangle}
\def\sandmm#1.#2.#3{%
\left\langle\smash{#1}{\vphantom1}^{-}\right|{\slash\!\!\! #2}%
\left|\smash{#3}{\vphantom1}^{-}\right\rangle}
\def\spab#1.#2.#3{\sandmm#1.#2.#3}
\def\spbb#1.#2.#3.#4{\sandpm#1.{\slash\!\!\! #2\slash\!\!\! #3}.#4}
\newbox\charbox
\newbox\slabox
\def\s#1{{      
        \setbox\charbox=\hbox{$#1$}
        \setbox\slabox=\hbox{$/$}
        \dimen\charbox=\ht\slabox
        \advance\dimen\charbox by -\dp\slabox
        \advance\dimen\charbox by -\ht\charbox
        \advance\dimen\charbox by \dp\charbox
        \divide\dimen\charbox by 2
        \raise-\dimen\charbox\hbox to \wd\charbox{\hss/\hss}
        \llap{$#1$}
}}
\def\ksl{\s{k}}

\def\beqa{\begin{eqnarray}}
\def\eeqa{\end{eqnarray}}
\def\beq{\begin{equation}}
\def\eeq{\end{equation}}
\def\hf{{\textstyle{\frac{1}{2}}}}
\def\ihf{{\textstyle{\frac{i}{2}}}}
\def\sst{\scriptscriptstyle}

\newcommand{\wh}[1]{\widehat{#1}}

\def\MHVb{$\overline{\text{MHV}}$}

\def\AB#1#2#3{\langle#1|#2|#3]}
\def\AA#1#2#3{\langle#1|#2|#3\rangle}

\def\A#1#2{\langle#1#2\rangle}
\def\B#1#2{[#1#2]}

\def\NP{N_{P}}
\def\NF{N_{F}}
\def\NN{$\mathcal{N}=4$}
\DeclareMathOperator{\F}{\mathit{F}}
\DeclareMathOperator{\Split}{\rm Split}
\DeclareMathOperator{\tr}{ {\rm tr}}
\def\trm{\tr_-}

\DeclareMathOperator{\Tri}{ {\rm F}^{1m}_3}
\DeclareMathOperator{\Ftme}{ {\rm F}^{2me}_4}
\DeclareMathOperator{\Fom}{ {\rm F}^{1m}_4}
\def\FF#1{\sideset{_2}{_1}\F\bigg(#1\bigg)}

\def\mc#1{\mathcal{#1}}
\preprint{
  IPPP/07/12\\
  SACLAY-SPHT-T07/044\\
  April, 2007}

\title{One-loop $\phi$-MHV amplitudes using the unitarity bootstrap}

\author{S. D. Badger$^*$,
    \ E. W. N. Glover$^\dagger$,
    \ Kasper Risager$^\ddagger$
	\\
	$^*$Service de Physique Theorique, CEA/Saclay, 91191 Gif-sur-Yvette, France
	\\
        $^\dagger$Department of Physics, University of Durham, Durham, DH1 3LE, UK 
	\\
	$^\ddagger$Niels Bohr Institute, Blegdamsvej 17, DK-2100, Copenhagen, Denmark
        \\
	E-mails: {simon.badger@cea.fr,
        e.w.n.glover@durham.ac.uk, risager@nbi.dk}.
	
    }

\abstract{

We consider a Higgs boson coupled to gluons via the five-dimensional effective
operator $H {\rm tr} G_{\mu\nu}G^{\mu\nu}$ produced by considering the 
heavy top quark limit of the one-loop coupling of
Higgs and gluons in the standard model.
We treat $H$ as the real part of a complex field $\phi$ 
that couples to the selfdual gluon field strengths
and compute the one-loop corrections to amplitudes
involving $\phi$, two colour adjacent negative helicity gluons and 
an arbitrary number of positive helicity gluons - the so-called $\phi$-MHV amplitudes.
We use four-dimensional unitarity to construct the cut-containing contributions
and the recently developed recursion relations to obtain the rational contribution
for an arbitrary number of external gluons.
We solve the recursion relations and give explicit results for up to four external gluons.
These amplitudes are relevant for Higgs plus jet production via gluon fusion in the 
limit where the top quark mass is large compared to all other scales in the problem.
\\ 

\today
}


\begin{document}

\section{Introduction}

As the time for physics at CERN's Large Hadron Collider (LHC)  approaches there is a 
great need for precision calculations of Standard Model processes. The eagerly
anticipated new physics signals typically result in complex multiparticle final states that are
contaminated by Standard Model contributions. Isolating the signal therefore relies on
precise theoretical calculations of background rates. 

At present, the rates for such processes can be estimated  relatively easily at leading
order (LO) in parton level perturbation theory. A wide range of tools are
available, for example ALPGEN~\cite{Mangano:alpgenII,Mangano:alpgenIII}, 
 the COMPHEP package~\cite{Pukhov:comphepI,Boos:comphepII},
HELEC/PHEGAS~\cite{Kanaki:helac,Papadopoulos:phegas},
MADGRAPH~\cite{Stelzer:madgraph,Maltoni:madevent} and
SHERPA/AMEGIC++~\cite{Gleisberg:sherpa,Krauss:amegic} 
which compute the
amplitudes numerically\footnote{COMPHEP, MADGRAPH and SHERPA/AMEGIC++ evaluate sums of Feynman diagrams.  
However, both ALPGEN and HELAC/PHEGAS are based on off-shell recursion
relations~\cite{Berends:offshellrecur,Kosower:offshellrecur,Caravaglios:alpgenI,Draggiotis:offshellrecur}.} 
and provide a suitable phase space over which they can be
integrated. However, the predicted event rates suffer large uncertainties due to the
choice of the unphysical renormalisation and factorisation scales  so that the
calculated rate is only an ``order of magnitude" estimate.  In addition, there is a
rather poor mismatch between the ``single parton becomes a jet" approach used in LO
perturbation theory and the complicated multi-hadron jet observed in experiment. 

While these problems cannot be entirely solved within perturbation theory, the situation
can be improved by calculating the strong next-to-leading order (NLO) corrections.  The
uncertainty on the rate may reduced to the 10\%-30\% level by including  the
next-to-leading order (NLO) corrections, which  may themselves be large when new
channels are accessed. In addition, the additional parton radiated into the final state
allows a better modelling of the inter- and intra-jet energy flow as well as identifying
regions where large logarithms must be resummed. 

The ingredients necessary for computing the NLO correction to a $n$ particle process are
well known. First, one needs the tree-level contribution for the $n+1$ particle process
where an additional parton is radiated. Second, one needs the one-loop $n$ particle
matrix elements. Both terms are infrared (and usually ultraviolet) divergent and must be
carefully combined to yield an infrared and ultraviolet finite NLO prediction. 

The real emission contribution is relatively well under control and can easily be
automated~\cite{Caravaglios:alpgenI,Mangano:alpgenII,Mangano:alpgenIII,Gleisberg:sherpa,Pukhov:comphepI,Boos:comphepII,Stelzer:madgraph,Maltoni:madevent}. The infrared singularities that
occur when a parton is soft (or when two partons become collinear) can then be removed
using well established (dimensional regularisation) techniques so that the ``subtracted"
matrix element is finite and can be evaluated in 4-dimensions~\cite{Catani:dipoles}. 

The bottleneck in deriving NLO corrections for multiparticle processes is computing the
one-loop amplitudes.  The standard approach is to compute the relevant Feynman diagrams
using a variety (or combination) of numerical and algebraic techniques. Much progress
has been made in this way, numerical evaluations of processes with up to six particles
have been performed~\cite{Denner:ee4f1,Denner:ee4f2,Ellis:1l6gnumer}. However,  one always observes large cancellations
between the contributions of different Feynman diagrams and the result is generally far
more compact than would naively be expected.  This is a strong hint that more direct
and  efficient ways of performing the calculation exist.  The simplicity can be realised
using  on-shell methods, where the cancellations due to gauge invariance and momentum
conservation are already present, to compute the amplitude.

In a series of pioneering papers,  Bern et al~\cite{BDDK:uni1,BDDK:uni2} developed the use of on-shell
methods at loop level by sewing together four-dimensional tree-level amplitudes and
using unitarity to reconstruct the (poly)logarithmic cut constructible part of the
amplitude.  A key feature of the unitarity method is that the   tree-level helicity
amplitudes are often very simple. However, the rational terms, that are produced when
computing non-supersymmetric amplitudes in $D=4-2\epsilon$ dimensions, were difficult to
obtain.

Recently, on-shell methods have received renewed attention with Witten's proposal of a
duality between \NN~supersymmetric Yang-Mills and a topological string theory
\cite{Witten:twstr}. New analytical methods, the MHV rules \cite{Cachazo:MHVtree} and
on-shell recursion relations \cite{Britto:rec}, have been derived and  have been
extremely successful in computing tree-level helicity amplitudes. Both of these
techniques can be proved using simple complex analysis and a knowledge of the
factorisation properties of the amplitudes on multi-particle poles
\cite{Britto:proof,Risager:mhvproof}.   Although developed for in the context of 
multigluon amplitudes, they have had wide ranging applications  including processes with
massive coloured scalars \cite{Badger:massrec,Forde:masssclr}, fermions
\cite{Badger:massvec,Schwinn:allborn,Schwinn:2006ca,Ferrario:2006np}, Higgs
bosons~\cite{Dixon:MHVhiggs,Badger:MHVhiggs2} and  vector bosons
\cite{Bern:EWcurrents,Badger:massvec}.  
A more complete set of references can be found in refs. ~\cite{Cachazo:twstrreview,Bern:onshellrev}.

Some of these on-shell ideas have been applied to one-loop amplitudes including the
application of maximally-helicity-violating (MHV) vertices~\cite{Brandhuber:n4},   a more
generalised unitarity~\cite{Britto:genuni} using complex momenta and the use of the
holomorphic anomaly~\cite{Britto:holo}.  Together with previous
results~\cite{BDDK:uni1,BDDK:uni2}, these new methods have led to the complete analytic
expressions for the cut-containing
contributions~\cite{Bedford:nonsusy,Britto:sqcd,Britto:ccqcd}  for one-loop QCD six-gluon
amplitude. Very recently an efficient method for direct extraction of the integral coefficients
exploiting complex momenta has also been proposed \cite{Forde:intcoeffs}.

However, these techniques are also intrinsically 4-dimensional and, although providing
effective ways of computing the cut-constructible (poly)logarithmic part of the amplitude,
suffer from the same limitations  as the older unitarity based methods and miss the rational
part. These ``missing" rational terms can be obtained directly from the unitarity approach
by taking the cut loop momentum to be in $D = 4 -2\epsilon$
dimensions~\cite{vanNeerven:dimreg} and there have been recent developments in this
direction~\cite{Anastasiou:DuniI,Mastrolia:2006ki,Britto:Duni,Anastasiou:DuniII}.

On the other hand,  the rational part is essentially tree-level-like  in containing only
poles in the complex plane.  One can therefore attempt to isolate these terms using an
on-shell recursion relation in an analogous way to tree level amplitudes. This is the 
``unitarity on-shell bootstrap" technique  which combines unitarity with on-shell
recursion~\cite{BDK:1lonshell,BDK:1lrecfin,Bern:bootstrap}. Other Feynman diagram based
methods have also been developed for calculating the rational terms
directly~\cite{Xiao:rationalI, Su:rationalII,Xiao:rationalIII,Ossola:rational,Binoth:rational}. Altogether,
it has already been possible to calculate the full set of complete 6 gluon helicity
amplitudes as well as closed forms for $n$-point amplitudes for specific 
helicities~\cite{Forde:MHVqcd,Berger:genhels,Berger:allmhv}, and the 
complete set of six photon amplitudes~\cite{Binoth:sixphoton,Ossola:sixphoton}.

As mentioned earlier, the MHV rules have also been successfully applied to tree  level QCD 
amplitudes involving a massive colourless scalar, the Higgs boson, in the large top mass
limit~\cite{Dixon:MHVhiggs,Badger:MHVhiggs2}.  This is achieved using a decomposition of the
Higgs field into its selfdual and anti-selfdual components, $\phi$ and $\phi^\dagger$. The
purpose of this paper is to apply the new on-shell methods to one-loop calculations of the
Higgs plus $n$-gluon amplitudes. Amplitudes of this kind have been considered in the context
of on-shell recursion in relations by Berger, Del Duca and
Dixon~\cite{Berger:higgsrecfinite} for the finite helicity configurations that vanish at
tree level, the $\phi$-``all plus" and $\phi$-``almost all plus" amplitudes. The
cut-constructible parts of the infrared divergent amplitudes involving the  $\phi$ and an
arbitrary number of negative helicity gluons were studied in ref.~\cite{Badger:1lhiggsallm}.
Here we consider the amplitudes with two adjacent negative helicity gluons and any number of
positive helicity gluons, the $\phi$-MHV amplitudes. These represent one of the ingredients
needed for the complete set of helicity amplitudes for Higgs to four gluon process which has
been computed using numerical methods in reference~\cite{Ellis:1lh24}.

Following~\cite{Bern:bootstrap}, we will split the calculation into two parts, 
evaluating the ``pure''
4-dimensional cut-constructible ${C}_n$ and rational ${R}_n$ 
parts of the leading colour contribution to
the one-loop $n$-point amplitude separately:
\begin{align}
	A_n^{(1)} &=  {C}_n+ {R}_n.
\end{align}
For simplicity, 
we have dropped the leading colour subscript, 
$A^{(1)}_{n;1}\equiv A^{(1)}_{n}$.
${C}_n$ contains all of the terms originating in box, triangle bubble loop integrals 
which are related to cut-containing logarithms and dilogarithms (or $\pi^2$) 
and which
are cut-constructible in 4-dimensions.  
In this paper we choose to
compute the cut-containing contribution
using the 1-loop MHV rules developed by Brandhuber, Spence and Travaglini~\cite{Brandhuber:n4}. 
In addition, ${C}_n$ contains fake
singularities that come from tensor loop integrals.   
To explicitly remove these spurious
singularities, it is convenient to introduce a cut-completing rational term,
so that the ``full'' cut-constructible term ${C}_n$ is given by
\begin{align}
	\wh{C}_n &= {C}_n+{CR}_n, 
\end{align}
with the corresponding modification to the rational part,
\begin{align}
	\wh{R}_n &= {R}_n-{CR}_n.
\end{align}
Because the rational part contains only simple poles,   the aim is to construct
this recursively using the multiparticle factorisation properties of
amplitudes.   This means constructing a direct recursive term, $R^D_n$, by
summing over products of lower point tree and one-loop amplitudes. By
construction, $R^D_n$ encodes the complete residues on the physical poles.  
The cut-completion contribution ${CR}_n$ may also give a contribution in the
physical channels which would then lead to double counting.
These potential unwanted contributions are removed by
the overlap term, so that 
\begin{align}
	\wh{R}_n &= {R}^D_n+{O}_n,
\end{align}
and
the full amplitude is given by
\begin{align}
	A_n^{(1)} &=  {C}_n + {CR}_n + {R}^D_n + {O}_n.
\end{align}
 
Our paper is organised as follows.  In Sec.~\ref{sec:higgs} we introduce the complex scalar
field $\phi$ and review the effective
interaction that couples it directly to gluons.   The relationship between 
$\phi$ (and
$\phi^\dagger$) and Higgs amplitudes is spelt out, along with out conventions for colour
ordering amplitudes and a selection explicit results for tree-level $\phi$-amplitudes.
We address the computation of the cut-constructible ${C}_n$ and cut-completion ${CR}_n$
parts of the one-loop amplitude in
section~\ref{sec:cut}.  The rational contributions are
considered in section~\ref{sec:rat} where
we establish the on-shell recursion relation ${R}^D_n$ and
give expressions for the overlap terms ${O}_n$.
We solve the arbitrary multiplicity results given in sections~\ref{sec:cut} and
\ref{sec:rat} for
the special case $n=4$ in
section~\ref{sec:four} and give an explicit analytic expression for the $A_4^{(1)}$ amplitude.
Section~\ref{sec:checks} is dedicated to a series of checks of our results.  We show
that our results have the correct infrared pole structure, satisfy the correct collinear
limits and we study the limit where the momentum of the Higgs boson becomes soft.
Finally, our findings are summarised in section~\ref{sec:conclusions}.  Two appendices are
enclosed that define our spinor notation and list the relevant one-loop functions that appear
in $C_n$.
 
\section{The Higgs Model}
\label{sec:higgs}

In the Standard Model the Higgs boson couples to gluons through
a fermion loop. The dominant contribution is from the top quark.
For large $m_t$, the top quark can be integrated out leading
to the effective interaction~\cite{Shifman:HggOperator2,Wilczek:HggOperator5},
\begin{equation}
 {\cal L}_{\sst H}^{\rm int} =
  \frac{C}{2}\, H \tr G_{\mu\nu}\, G^{\mu\nu}  \ .
 \label{HGGeff}
 \end{equation}
In the Standard Model,  
the strength of the interaction 
$C$ has been calculated up to order ${\cal{O}}(\alpha_{s}^{4})$
in~\cite{Chetyrkin:heffalpha3}. However, for our purposes we need it only up
to order ${\cal{O}}(\alpha_{s}^{2})$~\cite{Inami:Heff2l},
\begin{equation}
\label{eq:C}
C=\frac{\alpha_{s}}{6\pi v}\left(1+\frac{11}{4}\frac{\alpha_{s}}{\pi}+
\ldots\right)\, ,
\end{equation}
with $v = 246$~GeV.
This approximation works very well under the condition that the kinematic
scales involved are smaller
than twice the top quark mass $M_{t}$.~\cite{Kramer:1996iq,Baur:ptdist,Ellis:ptdist}

The MHV structure of the Higgs-plus-gluons amplitudes is best
elucidated \cite{Dixon:MHVhiggs} by considering $H$ to be the real
part of a complex field $\phi = \frac{1}{2}( H + i A )$, so that
\begin{eqnarray}
 {\cal L}^{\rm int}_{H,A} &=&
\frac{C}{2} \Bigl[ H \tr G_{\mu\nu}\, G^{\mu\nu}
             + i A \tr G_{\mu\nu}\, {}^*G^{\mu\nu} \Bigr]
 \label{effinta}\nonumber \\
\lefteqn{=
C \Bigl[ \phi \tr G_{{\sst SD}\,\mu\nu}\, G_{\sst SD}^{\mu\nu}
 + \phi^\dagger \tr G_{{\sst ASD}\,\mu\nu} \,G_{\sst ASD}^{\mu\nu} \Bigr]
 }\nonumber \\
 \label{effintb}
\end{eqnarray}
where the purely selfdual (SD) and purely anti-selfdual (ASD)
gluon field strengths are given by
$$
G_{\sst SD}^{\mu\nu} = \hf(G^{\mu\nu}+{}^*G^{\mu\nu}) \ , \quad
G_{\sst ASD}^{\mu\nu} = \hf(G^{\mu\nu}-{}^*G^{\mu\nu}) \ , 
$$
with
$$
{}^*G^{\mu\nu} \equiv \ihf \epsilon^{\mu\nu\rho\sigma} G_{\rho\sigma} \ .
$$
The important observation of \cite{Dixon:MHVhiggs} was that, due to selfduality, the amplitudes for
$\phi$ plus $n$ gluons, and those for $\phi^\dagger$ plus $n$ gluons,
each have a simpler structure than the gluonic amplitudes for
either $H$ or $A$. Amplitudes can be constructed for
$\phi$ plus $n$ gluons  
and for $\phi^\dagger$ plus $n$ gluons separately.

The decomposition of the $HGG$ and $AGG$ vertices into the self-dual and the anti-self-dual
terms \eqn{effintb} means that the  Higgs and pseudoscalar Higgs amplitudes are obtained from
$\phi$ and $\phi^\dagger$ amplitudes.   
\begin{eqnarray}
\label{eq:H}
 {\cal A}^{(m)}_n(H,g_1^{\lambda_1},\ldots,g_n^{\lambda_n}) &=&
 {\cal A}^{(m)}_n(\phi,g_1^{\lambda_1},\ldots,g_n^{\lambda_n}) + 
\label{eq:A}
 {\cal A}^{(m)}_n(\phi^\dagger,g_1^{\lambda_1},\ldots,g_n^{\lambda_n}) , \\
 {\cal A}^{(m)}_n(A,g_1^{\lambda_1},\ldots,g_n^{\lambda_n}) &=&
 {\cal A}^{(m)}_n(\phi,g_1^{\lambda_1},\ldots,g_n^{\lambda_n}) - 
{\cal A}^{(m)}_n(\phi^\dagger,g_1^{\lambda_1},\ldots,g_n^{\lambda_n}) .
\end{eqnarray}
However, parity further relates $\phi$ and $\phi^\dagger$ 
amplitudes,
\begin{align}
\label{eq:phidagger}
 {\cal A}^{(m)}_n(\phi^\dagger,g_1^{\lambda_1},\ldots,g_n^{\lambda_n}) &= 
 \left({\cal A}^{(m)}_n(\phi,g_1^{-\lambda_1},\ldots,g_n^{-\lambda_n}) \right)^*.
\end{align}
From now on, we will only consider $\phi$ amplitudes, knowing that all others can be obtained
using eqs.~(\ref{eq:H})--(\ref{eq:phidagger}).

\subsection{Colour ordering}

The tree level amplitudes for a $\phi$ and $n$ gluons 
can be decomposed into colour ordered amplitudes as~\cite{Dawson:Htomultijet,DelDuca:Hto3jets},
\begin{align}
	{\cal A}^{(0)}_n(\phi,\{k_i,\lambda_i,a_i\}) = 
	i C g^{n-2}
	\sum_{\sigma \in S_n/Z_n}
	\tr(T^{a_{\sigma(1)}}\cdots T^{a_{\sigma(n)}})\,
	A^{(0)}_n(\phi,\sigma(1^{\lambda_1},..,n^{\lambda_n})).
	\label{TreeColorDecompositionQ}
\end{align}
Here $S_n/Z_n$ is the group of non-cyclic permutations on $n$
symbols, and $j^{\lambda_j}$ labels the momentum $p_j$ and helicity
$\lambda_j$ of the $j^{\rm th}$ gluon, which carries the adjoint
representation index $a_i$.  The $T^{a_i}$ are fundamental
representation SU$(N_c)$ color matrices, normalized so that
${\rm Tr}(T^a T^b) = \delta^{ab}$.  The strong coupling constant is
$\alpha_s=g^2/(4\pi)$.

Tree-level amplitudes with a single quark-antiquark pair 
can be decomposed into colour-ordered amplitudes as follows,
\begin{eqnarray}
\lefteqn{
{\cal A}_n(\phi,\{p_i,\lambda_i,a_i\},\{p_j,\lambda_j,i_j\}) }\\
&&= 
i C g^{n-2}
\sum_{\sigma \in S_{n-2}} (T^{a_{\sigma(2)}}\cdots T^{a_{\sigma(n-1)}})_{i_1i_n}\,
A_n(\phi,1^{\lambda},\sigma(2^{\lambda_2},\ldots,{(n-1)}^{\lambda_{n-1}}),
n^{-\lambda})\,.\nonumber 
\label{TreeColorDecomposition}
\end{eqnarray}
where $S_{n-2}$ is the set of permutations of $(n-2)$ gluons.
Quarks are characterised with fundamental colour 
label $i_j$  and
helicity $\lambda_j$ for $j=1,n$.
By current conservation, the quark and antiquark helicities are  related such
that $\lambda_1 = -\lambda_n \equiv \lambda$ where $\lambda = \pm \hf$.

The l-loop gluonic amplitudes which are the main subject of this paper
follow the same colour ordering as the pure QCD amplitudes \cite{BDDK:uni1}
and can be decomposed as \cite{Berger:higgsrecfinite,Badger:1lhiggsallm},
\begin{align}
	\mc A^{(1)}_n(\phi,\{k_i,\lambda_i,a_i\}) &= i C g^{n}
	\sum_{c=1}^{[n/2]+1}\sum_{\sigma \in S_n/S_{n;c}} G_{n;c}(\sigma)
	A^{(1)}_n(\phi,\sigma(1^{\lambda_1},\ldots,n^{\lambda_n}))
	\label{eq:1lhtogcolour}
\end{align}
where
\begin{align}
	G_{n;1}(1) &= N \tr( T^{a_1}\cdots T^{a_n} ) \\
	G_{n;c}(1) &=   \tr( T^{a_1}\cdots T^{a_{c-1} } )
			\tr( T^{a_c}\cdots T^{a_n} )
	\,\, , \, c>2.
	\label{eq:colourfactors}
\end{align}
The sub-leading terms can be computed by summing over various permutations of the leading colour
amplitudes \cite{BDDK:uni1}.

\subsection{Tree level $\phi$ amplitudes}

As noted in~\cite{Dixon:MHVhiggs} the all-plus and almost all-plus $\phi$ amplitudes
vanish,
\begin{align}
 A^{(0)}_n(\phi,g_1^+,g_2^+,g_3^+,\ldots,g_n^+) &= 0 \, , \\
  A^{(0)}_n(\phi,g_1^-,g_2^+,g_3^+,\ldots,g_n^+) &= 0 \, ,
\end{align}
for all $n$.

The tree $\phi$-amplitudes, with precisely two negative helicities
are the
first non-vanishing $\phi$ amplitudes. These amplitudes are
the $\phi$-MHV amplitudes and
general factorisation properties now imply that they have to be extremely
simple~\cite{Dixon:MHVhiggs}. For the case when legs $q$ and $p$ have negative helicity,
they are given by 
\begin{equation}
 A^{(0)}_n(\phi,g_1^+,g_2^+,\ldots, g_p^-, \ldots, g_q^-, \ldots ,g_n^+)   
=
 \frac{{\spa{p}.{q}}^4} { \spa1.2 \spa2.3 \cdots \spa{n-1,}.{n} \spa{n}.1  },
\label{eq:phi-mhv}
\end{equation}
 In fact, the expression
\eqn{eq:phi-mhv} for the $\phi$-MHV $n$-gluon amplitude has precisely
the same form
as the MHV $n$-gluon amplitudes in pure QCD \cite{Parke:ngluon}. The only difference is that the total momentum carried by gluons,
$p_1+p_2+\ldots+p_n=-p_{\phi}$ is the momentum
carried by the $\phi$-field and is non-zero. 

There are two $\phi$-MHV amplitudes involving a quark pair,
 \begin{eqnarray}
	A_n(\phi,q^{-}_1,\ldots,g^-_r,\ldots,\bar{q}^{+}_n) 
	&=& \frac{\spa{r}.{1}^3\spa{r}.{n}}{\spa1.2 \spa2.3 \cdots \spa{n-1,}.{n} \spa{n}.1 } \label{eq:2qMHV1},\\
	A_n(\phi,q^{+}_1,\ldots,g^-_r,\ldots,\bar{q}^{-}_n) &=& 
	\frac{\spa{r}.{1}\spa{r}.{n}^3}{\spa1.2 \spa2.3 \cdots \spa{n-1,}.{n} \spa{n}.1 }.
	\label{eq:2qMHV2}
\end{eqnarray}

We note in passing that the 
tree $\phi$-amplitude with all negative helicity gluons, the $\phi$-all-minus amplitude,
also has a simple structure~\cite{Dixon:MHVhiggs},
 \begin{align}
	\lefteqn{A^{(0)}_n(\phi;1^-,\ldots,n^-)} \nonumber\\
	&= \left(-1\right)^n
	\frac{m_H^4}{\spb1.2 \spb2.3 \cdots \spb{n-1,}.{n} \spb{n}.1}\ .
	\label{eq:allminus}
\end{align}

Amplitudes with fewer (but more than two) negative helicities have been 
computed with Feynman diagrams (up to 4 partons) in Ref.~\cite{DelDuca:Hto3jets}
and using MHV rules and on-shell recursion relations in Refs.~\cite{Dixon:MHVhiggs,Badger:MHVhiggs2}.

The main goal of this paper is the construction of the one-loop $\phi$-MHV amplitude
with two adjacent negative helicities.
For definiteness, we focus on the specific helicity configuration  $(1^-,2^-,3^+,\cdots,n^+)$.

\section{The cut-constructible contributions}
\label{sec:cut}
 
In a landmark paper, Brandhuber, Spence and Travaglini~\cite{Brandhuber:n4}  showed that it is
possible to calculate one-loop MHV amplitudes in \NN\ using MHV rules.  
The calculation has many similarities to the unitarity based approach of 
Refs.~\cite{BDDK:uni1,BDDK:uni2},  the main difference being
that the MHV rules reproduce the cut-constructible parts of the amplitude directly, 
without having to worry about double counting. This is the method that we wish to employ here.

\begin{figure}[b]
	\psfrag{L1}{$L_1$}
	\psfrag{L2}{$L_2$}
	\psfrag{A1}{$A_L$}
	\psfrag{A2}{$A_R$}
	\begin{center}
		\includegraphics[width=6cm]{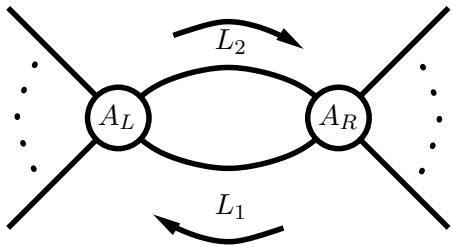}
	\end{center}
	\caption{A generic one loop MHV diagram or unitarity cut.}
	\label{fig:1lmhvdiag}
\end{figure}

The four-dimensional cut-constructible part of one-loop amplitudes can be constructed by joining two
on-shell  vertices by two scalar propagators, both of which need to be continued off-shell. A generic
diagram is shown in figure \ref{fig:1lmhvdiag} and the full amplitude will be a sum over all possible
permutations and helicity configurations.  
In the BST approach the propagators are
continued off-shell as for the tree level MHV rules and can be 
written \cite{Brandhuber:n4},
\begin{equation}
	L_i = l_i+z_i\eta.
	\label{eq:loopoffshell}
\end{equation}
Loop integration over the $L_i$ are related to phase space integration over
$l_i$ and trivial integrals over $z_i$. 
Note that this approach actually performs all integrals and therefore
there is a one-to-one identification of cut diagrams and cut-constructible 
functions.   However, it relies on being able to perform the phase space integral.
For the cases considered in this paper, the phase space integrals are known and are
directly related to loop functions. 

A generic diagram can be written:
\begin{align}
	\mc D = \frac{1}{(2\pi)^4}\int \frac{d^4L_1}{L_1^2}\frac{d^4L_2}{L_2^2}
	\delta^{(4)}(L_1-L_2-P)A_L(l_1,-P,-l_2) A_R(l_2,P,-l_1)
\end{align}
where $A_{L(R)}$ are the amplitudes for the left(right) vertices and $P$ is the sum of momenta
incoming to the right hand amplitude. The important step is the
evaluation of this expression is to re-write the integration measure as an integral over the
on-shell degrees of freedom and a separate integral over the complex variable $z$
\cite{Brandhuber:n4}:
\begin{align}
	\frac{d^4L_1}{L_1^2}\frac{d^4L_2}{L_2^2} &=
	(4i)^2\frac{dz_1}{z_1}\frac{dz_2}{z_2}d^4l_1d^4l_2\delta^{(+)}(l_1^2)\delta^{(+)}(l_2^2)\nonumber\\
	&=(4i)^2 \frac{2 dzdz'}{(z-z')(z+z')}d^4l_1d^4l_2\delta^{(+)}(l_1^2)\delta^{(+)}(l_2^2),
	\label{eq:intmeasure}
\end{align}
where $z=z_1-z_2$ and $z'=z_1+z_2$. The integrand can only depend on $z,z'$ through the momentum
conserving delta function,
\begin{equation}
	\delta^{(4)}(L_1-L_2-P) = \delta^{(4)}(l_1-l_2-P+z\eta) = \delta^{(4)}(l_1-l_2-\wh{P}),
\end{equation}
where $\wh{P} = P-z\eta$. This means that the integral over $z'$ can be performed so that,
\begin{align}
	\mc D &=\frac{(4i)^2 2\pi i}{(2\pi)^4}\int\frac{dz}{z}\int d^4l_1 d^4l_2\delta^{(+)}(l_1^2)\delta^{(+)}(l_2^2)
	\delta^{(4)}(l_1-l_2-\wh{P}) A_L(l_1,-P,-l_2)  A_R(l_2,P,-l_1) \nonumber\\
	&=(4i)^2 2\pi i\int\frac{dz}{z}\int d{\rm LIPS}^{(4)}(-l_1,l_2,\wh{P}) A_L(l_1,-P,-l_2)
	 A_R(l_2,P,-l_1),
\end{align}
where,
\begin{equation}
	d{\rm LIPS}^{(4)}(-l_1,l_2,\wh{P}) = \frac{1}{(2\pi)^4} d^4l_1 d^4l_2\delta^{(+)}(l_1^2)\delta^{(+)}(l_2^2)\delta^{(4)}(l_1-l_2-\wh{P})	
\end{equation}
The phase space integral is regulated using dimensional regularisation. 
Tensor integrals arising from the product of tree amplitudes can be reduced 
to scalar integrals either by using spinor algebra
or standard Passarino-Veltman reduction. The remaining scalar integrals have been evaluated
previously by van Neerven~\cite{vanNeerven:dimreg}.

\subsection{Pure cut contributions}

\begin{figure}
	\psfrag{m}{\tiny$-$}
	\psfrag{mp}{\tiny$\mp$}
	\psfrag{pm}{\tiny$\pm$}
	\psfrag{p}{\tiny$+$}
	\psfrag{phi}{\small$\phi$}
	\psfrag{nm1}{\small$(n-1)^+$}
	\psfrag{im1}{\small$(i-1)^+$}
	\psfrag{ip1}{\small$(i+1)^+$}
	\psfrag{i}{\small$i^+$}
	\psfrag{n}{\small$n^+$}
	\psfrag{1}{\small$1^-$}
	\psfrag{2}{\small$2^-$}
	\psfrag{3}{\small$3^+$}
	\begin{center}
		\includegraphics[width=\textwidth]{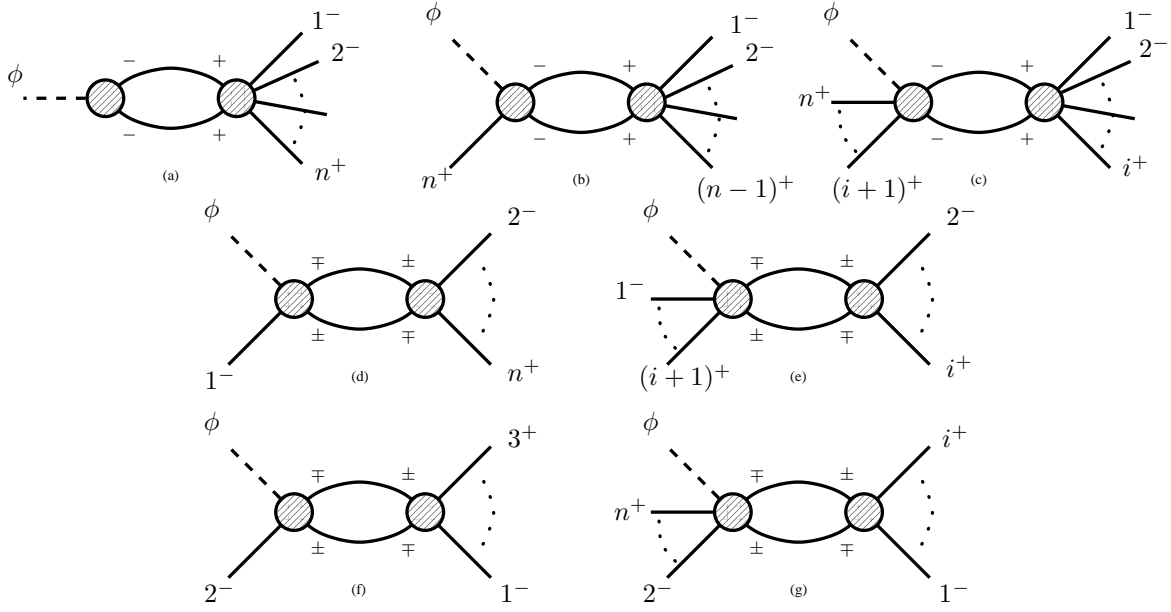}
	\end{center}
	\caption{The MHV loop diagrams contributing to the $\phi\to g_1^-g_2^-g_3^+\ldots g_n^+$
	amplitude.}
	\label{fig:hmm}
\end{figure}

The pure cut contribution is constructed 
by connecting two tree-level vertices and the 
seven independent topologies are shown in
Figure \ref{fig:hmm}. 
Note that the last four topologies helicity configurations allow
both fermionic and gluonic contributions. Note also that all fermion loops
always appear in association with a factor of $\NF$, the number of fermion species, and a factor of $-1$.

Let us consider diagram \ref{fig:hmm}(a) to begin with. We can take the momenta to be labelled
from 1 to $n$ around the right hand amplitude so that we consider a cut in the
$s_{1,n}$ channel, i.e. when the momentum flowing across the cut is $p_1+p_2+\ldots+p_n$ corresponding to
an invariant mass $s_{1,n}=(p_1+p_2+\ldots+p_n)^2$.\footnote{Note that when $j < i$, $s_{i,j}
\equiv (p_i+\ldots+p_n+p_1+\ldots p_j)^2$.}
Other diagrams
with this topology are accessible by permuting the arguments of the $s_{1,n}$ channel. The
product of the two vertices can be written:
\begin{align}
	A_L A_R &=
	-\frac{m_H^4}{\A{l_1}{l_2}^2}\frac{\A{1}{2}^4}{ \A{n}{l_1}\A{l_1}{l_2}\A{l_2}{1}
	\prod_{\alpha=1}^{n-1} \A{\alpha}{\alpha+1}}\nonumber\\
	&= A^{(0)}(\phi;1^-,2^-,\ldots,n^+)\frac{\A{l_1}{l_2}\A{n}{1}}{\A{l_2}{1}\A{n}{l_1}}.
\end{align}
Applying a Schouten identity to the numerator and using momentum conservation in the form $l_1 =
l_2 +\wh{P}_{1,n}$ we find,
\begin{align}
	A_L A_R &= A^{(0)}\left(
	-\frac{N(\wh{P}_{1,n},p_1,p_n)}{(l_1-p_n)^2(l_2+p_1)^2} - \frac{\wh{P}_{1,n}\cdot p_n}{(l_1-p_n)^2}
        + \frac{ \wh{P}_{1,n}\cdot p_1 }{(l_2+p_1)^2}\right),
\end{align}
where $N(P,p_1,p_2) = P^2(p_1\cdot p_2)-2(P\cdot p_1)(P\cdot p_2)$. This is now written in terms of
scalar integrals so we can directly use the results of van Neerven~\cite{vanNeerven:dimreg} to perform the phase space integration:
\begin{align}
	 \int d^D{\rm LIPS}&(-l_1,l_2,P)
		\frac{N(P,p_1,p_2)}{(l_1+p_1)^2(l_2+p_2)^2}
		=\nonumber\\
		&\frac{c_\Gamma}{(4\pi)^2\e^2}2i\sin(\pi\e) 
		\mu^{2\e}|P^2|^{-\e}\FF{1,-\e;1-\e; \frac{ p_1\cdot p_2 P^2 }{ N(P,p_1,p_2) } } \\
	 \int d^D{\rm LIPS}&(-l_1,l_2,P)
		\frac{2(P\cdot p_1)}{(l_1+p_1)^2}
		=\frac{c_\Gamma}{(4\pi)^2\e^2}2i\sin(\pi\e) 
		\mu^{2\e}|P^2|^{-\e} \\
	 \int d^D{\rm LIPS}&(-l_1,l_2,P)\phantom{\frac{2(P\cdot p_1)}{(l_1+p_1)^2}}
		=
		-\frac{c_\Gamma}{(4\pi)^2\e(1-2\e)}2i\sin(\pi\e) 
		\mu^{2\e}|P^2|^{-\e}
	\label{eq:dlips}
\end{align}
where the factor $c_\Gamma$ is given by,
\begin{equation}
	c_\Gamma = (4\pi)^{\e-2}\frac{\Gamma(1+\e)\Gamma^2(1-\e)}{\Gamma(1-2\e)}.
\end{equation}
The final integration is over the $z$ variable. However, the only dependence on $z$ appears through the
quantity $\wh{P}_{1,n}$\footnote{Through a suitable choice of 
$\eta$, one can always ensure that $N(P,p_1,p_2)$ is independent of $z$~\cite{Brandhuber:n4}}
 so it is convenient to make a change of variables,
\begin{equation}
	\frac{dz}{z} = \frac{d(\wh{P})^2}{P^2-\wh{P}^2}
\end{equation}
to produce a dispersion integral that will re-construct the parts of
the cut-constructible amplitude proportional to $(s_{1,n})^{-\e}$,
\begin{equation}
	\int \frac{d(\wh{P})^2}{P^2-\wh{P}^2} 2i\sin(\pi\e)|\wh{P}^2|^{-\e} = 2\pi i (-P^2)^{\e}.
	\label{eq:dispersion}
\end{equation}
The final result for this diagram then reads:
\begin{equation}
	\mc D^{1,n} = \frac{c_\Gamma}{\e^2} A^{(0)}\left( \frac{\mu^2}{-s_{1,n}} \right)^{\e}
	\left( \FF{1,-\e;1-\e; \frac{ p_1\cdot p_n s_{1,n} }{ N(P_{1,n},p_1,p_n) } } + 1  \right).
	\label{eq:1nch}
\end{equation}

The other ``gluon-only" channels (Figs.~\ref{fig:hmm}(b) and (c))
reduce to scalar integrals in the same way and we merely quote the results,
\begin{align}
	\mc D^{1,n-1} &= \frac{c_\Gamma}{\e^2} A^{(0)}\left( \frac{\mu^2}{-s_{1,n-1}} \right)^{\e}
	\Bigg( 
	\FF{1,-\e;1-\e; \frac{ p_1\cdot p_n s_{1,n-1} }{ N(P_{1,n-1},p_1,p_n) } }\nonumber\\
	&+\FF{1,-\e;1-\e; \frac{ p_n\cdot p_{n-1} s_{1,n-1} }{ N(P_{1,n-1},p_n,p_{n-1}) } }\nonumber\\
	&-\FF{1,-\e;1-\e; \frac{ p_1\cdot p_{n-1} s_{1,n-1} }{ N(P_{1,n-1},p_1,p_{n-1}) } }
	+ 1  \Bigg) \label{eq:1nm1ch} \\
	\mc D^{1,i} &= \frac{c_\Gamma}{\e^2} A^{(0)}\left( \frac{\mu^2}{-s_{1,i}} \right)^{\e}
	\Bigg( 
	\FF{1,-\e;1-\e; \frac{ p_1\cdot p_{i+1} s_{1,i} }{ N(P_{1,i},p_1,p_{i+1}) } }\nonumber\\
	&+\FF{1,-\e;1-\e; \frac{ p_n\cdot p_i s_{1,i} }{ N(P_{1,i},p_n,p_i) } }\nonumber\\
	&-\FF{1,-\e;1-\e; \frac{ p_1\cdot p_i s_{1,i} }{ N(P_{1,i},p_1,p_i) } }\nonumber\\
	&-\FF{1,-\e;1-\e; \frac{ p_n\cdot p_{i+1} s_{1,i} }{ N(P_{1,i},p_n,p_{i+1}) } }
	\Bigg).
	\label{eq:1ich}
\end{align}
The arguments of these expressions can be straightforwardly 
permuted to produce results for channels \ref{fig:hmm}(a),(b) and (c) for all gluon configurations.

\begin{figure}[t]
	\psfrag{m1}{\tiny$1^-$}
	\psfrag{m2}{\tiny$2^-$}
	\psfrag{n}{\tiny$n^+$}
	\psfrag{phi}{\tiny$\phi$}
	\psfrag{XXX}{\Large$\left( 1-\frac{\NF}{N} \right) 
	\left( \frac{2\trm(2Pn1)^3}{3 s_{12}^3(2P.p_n)^3}-\frac{\trm(2Pn1)^2}{
	s_{12}^2(2P.p_n)^2}\right)+4\left(1-\frac{\NF}{4N}\right)\frac{\trm(2Pn1)}{s_{12}(2P.p_n)}$}
	\begin{center}
		\includegraphics[width=1.05\textwidth]{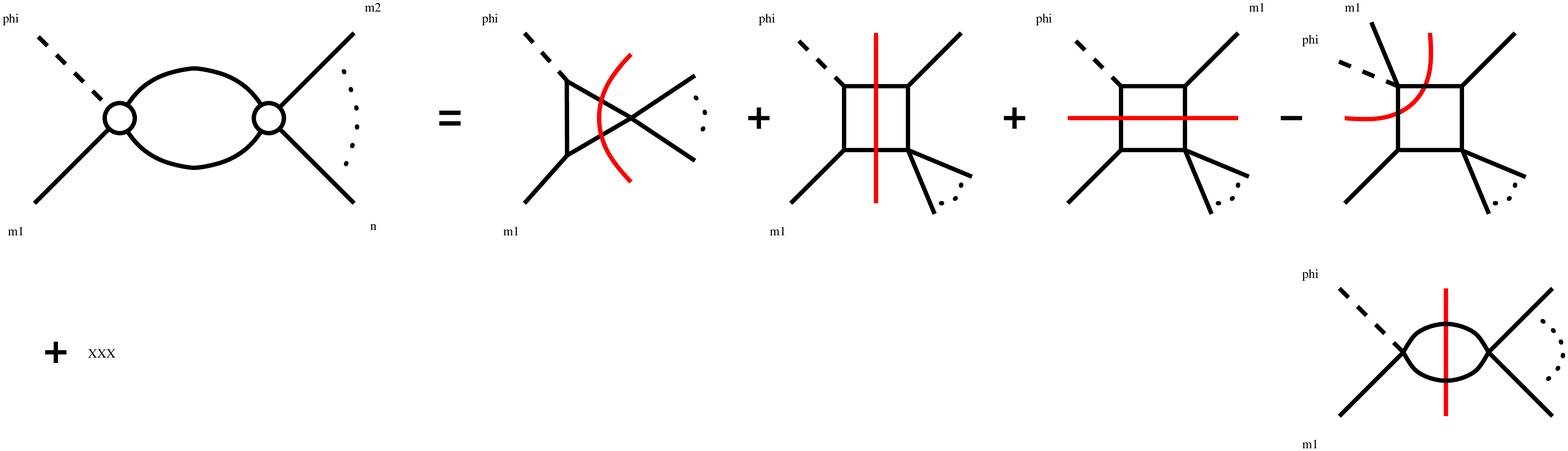}
	\end{center}
	\caption{Decomposition of the MHV diagram of fig. 2(d) contributing to the $P_{2,n}$ channel}
	\label{fig:2nch}
\end{figure}

\begin{figure}[t]
	\psfrag{m1}{\tiny$1^-$}
	\psfrag{m2}{\tiny$2^-$}
	\psfrag{i}{\tiny$i^+$}
	\psfrag{ip1}{\tiny$(i+1)^+$}
	\psfrag{phi}{\tiny$\phi$}
	\psfrag{XXX}{\Large$\left( 1-\frac{\NF}{N} \right) 
	\left( \frac{2\trm(2Pi1)^3}{3 s_{12}^3(2P.p_i)^3}-\frac{\trm(2Pi1)^2}{
	s_{12}^2(2P.p_i)^2}\right)+4\left(1-\frac{\NF}{4N}\right)\frac{\trm(2Pi1)}{s_{12}(2P.p_i)}$}
	\psfrag{YYY}{\Large$\left( 1-\frac{\NF}{N} \right) 
	\left( \frac{2\trm(2P(i+1)1)^3}{3 s_{12}^3(2P.p_{i+1})^3}-\frac{\trm(2P(i+1)1)^2}{
	s_{12}^2(2P.p_{i+1})^2}\right)+4\left(1-\frac{\NF}{4N}\right)\frac{\trm(2P(i+1)1)}{s_{12}(2P.p_{i+1})}$}
	\begin{center}
		\includegraphics[width=1.05\textwidth]{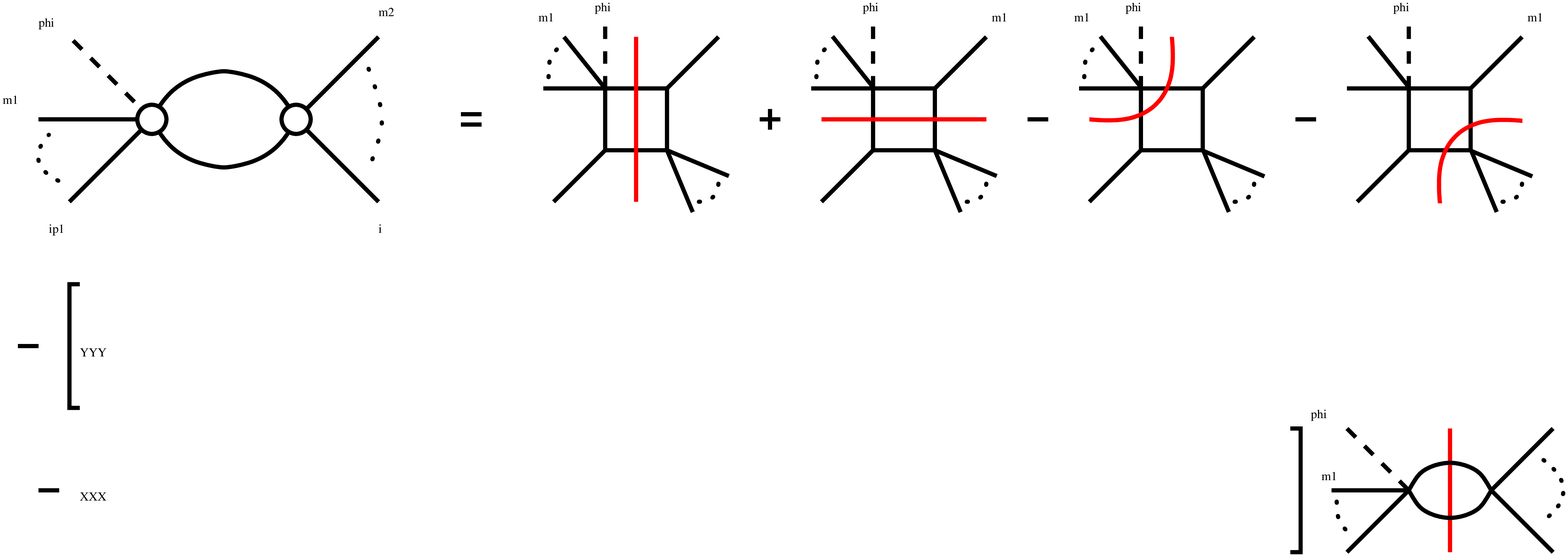}
	\end{center}
	\caption{Decomposition of the MHV diagram of fig. 2(e) contributing to the $P_{2,i}$ channel}
	\label{fig:2ich}
\end{figure}

When considering the channels with alternating helicity configurations around the loop we find that
even after the Schouten
identities have been applied, we are still left with tensor integrals which must be further reduced to
scalar integrals by expanding in terms of all possible tensor structures. This feature has also been
seen in the context of finding the cut-constructible part of pure QCD amplitudes and was also
addressed applying Passarino-Veltman reduction \cite{Bedford:nonsusy}.
For diagram \ref{fig:hmm}(d), the $s_{2,n}$ channel, the presence of tensor integrals and fermion
loops results in new structures of order $1/\e$.
The result for this diagram is,
\begin{align}
	\mc D^{2,n} = \frac{c_\Gamma}{\e^2} A^{(0)}\left( \frac{\mu^2}{-s_{2,n}} \right)^{\e}
	&\Bigg[ 
	1+\FF{1,-\e;1-\e; \frac{ p_2\cdot p_n s_{2,n} }{ N(P_{2,n},p_2,p_n) } }\nonumber\\
	&+\FF{1,-\e;1-\e; \frac{ p_1\cdot p_n s_{2,n} }{ N(P_{2,n},p_1,p_n) } }\nonumber\\
	&-\FF{1,-\e;1-\e; \frac{ p_1\cdot p_2 s_{2,n} }{ N(P_{2,n},p_1,p_2) } }\nonumber\\
	&- \left(1-\frac{\NF}{N}\right)\left( \frac{2\trm(2P_{3,n-1}n1)^3}{3s_{12}^3(2P\cdot n)^3} 
	+ \frac{\trm(P_{3,n-1}n1)^2}{s_{12}^2(2P\cdot n)^2} \right)\frac{\e}{1-2\e}\nonumber\\ 
	&- 4\left(1-\frac{\NF}{4N}\right)\left( \frac{\trm(P_{3,n-1}n1)}{s_{12}(2P\cdot n)}
	\right) \frac{\e}{1-2\e} 
	\Bigg],
	\label{eq:2nch}
\end{align}
where we have introduced the shorthand notation,
\begin{equation}
\trm(abcd) = \spa a.b \spb b.c \spa c.d \spb d.a  .
\end{equation}
This is better illustrated in figure \ref{fig:2nch} which shows the cuts of
each integral function that appear. Figure \ref{fig:2ich} shows the decomposition of the $s_{2,i}$
channels (figure \ref{fig:hmm}(e)) which follows exactly the same steps as the previous case.
Diagrams \ref{fig:hmm}(f) and \ref{fig:hmm}(g) are
analogous to diagrams \ref{fig:hmm}(d) and \ref{fig:hmm}(e) and can be found be permuting the
arguments: $1,2,\ldots,n \to 2,1,n,\ldots,3$.

When summing over all the possible diagrams we find that the bubble integrals always appear in the
combination,
\begin{equation}
	{\rm Bub}(s)-{\rm Bub}(t) = \mc O(\e^0).
\end{equation}
Hence all of the $1/\epsilon$ poles coming from bubble functions vanish
leaving the expected combination of 
boxes and triangles proprotional to the tree amplitude\cite{Catani:irpoles,Giele:irpoles}.
The combination of bubble
integrals can be written in terms of a basis of pure logarithms,
\begin{equation}
	L_k(s,t) = \frac{\log(s/t)}{(s-t)^k}.
	\label{eq:finitelogs}
\end{equation}
These logarithmic contributions are not proportional to the tree amplitude, but are multiplied by new
spinor structures written in terms of traces. The full, unrenormalised result for 
this specific MHV helicity configuration for $n \ge 3$ is thus:
\begin{align}
	&C_n(\phi,1^-,2^-,3^+\ldots,n^+) = 
	 c_\Gamma A^{(0)}_n(\phi,1^-,2^-,3^+,\ldots,n^+)
	\Bigg[\sum_{i=1}^n \left(\Tri(s_{i,n+i-2}) - \Tri(s_{i,n+i-1})\right)\nonumber\\
	&-\frac{1}{2}\sum_{i=1}^n\sum_{j=i+3}^{n+i-1}\Ftme(s_{i,j},s_{i+1,j-1};s_{i,j+1},s_{i+1,j})
	-\frac{1}{2}\sum_{i=1}^n\Fom(s_{i,i+2};s_{i,i+1},s_{i+1,i+2})\nonumber\\
	&+\sum_{i=4}^{n}\bigg(
	\frac{\NP}{3}\frac{\trm(1P_{i,n}(i-1)2)^3}{s_{12}^3}L_3(s_{i-1,1},s_{i,1})
	+\frac{\NP}{3}\frac{\trm(2P_{3,i-1}i1)^3}{s_{12}^3}L_3(s_{2,i},s_{2,i-1})
	\nonumber\\
	&\phantom{\sum_{i=4}^{n}\bigg(}
	-\frac{\NP}{2}\frac{\trm(1P_{i,n}(i-1)2)^2}{s_{12}^2}L_2(s_{i-1,1},s_{i,1})
	-\frac{\NP}{2}\frac{\trm(2P_{3,i-1}i1)^2}{s_{12}^2}L_2(s_{2,i},s_{2,i-1})
	\nonumber \\
	&\phantom{\sum_{i=4}^{n}\bigg(}
	+\frac{\NP}{6}\frac{\trm(1P_{i,n}(i-1)2)}{s_{12}}L_1(s_{i-1,1},s_{i,1})
	+\frac{\NP}{6}\frac{\trm(2P_{3,i-1}i1)}{s_{12}}L_1(s_{2,i},s_{2,i-1})
	\nonumber \\
	&\phantom{\sum_{i=4}^{n}\bigg(}
	+\frac{\beta_0}{N}\,\frac{\trm(1P_{i,n}(i-1)2)}{s_{12}}L_1(s_{i-1,1},s_{i,1})
	+\frac{\beta_0}{N}\,\frac{\trm(2P_{3,i-1}i1)}{s_{12}}L_1(s_{2,i},s_{2,i-1})
	\bigg)
	\Bigg],
	\label{eq:1lhmmCCa}
\end{align}
where the one-mass triangle $\Tri$  and box functions ${\rm F}_4$ are defined in
Appendix~\ref{app:scalarintegrals}.  For convenience, we have introduced
\begin{equation}
\beta_0 = \frac{11N-2\NF}{3}, \qquad\qquad\NP= 2\left(1-\frac{\NF}{N}\right).
\end{equation}
Note also that summations of the form 
$\sum_a^b$ are understood to vanish when $b < a$. If we add this term together with its complex
conjugate with the appropriate momentum re-labeling then we find agreement with the known
Higgs MHV amplitude in the case of $n=3$~\cite{Schmidt:1lHto3jets}.

Eq.~\eqref{eq:1lhmmCCa} can be rewritten in a form which is 
more convenient when computing the completion and overlap terms, namely
\begin{align} 
C_n(\phi,1^-,2^-,3^+\ldots,n^+) =&\nonumber \\
	 c_\Gamma A^{(0)}_n(\phi,1^-,2^-,3^+,\ldots,n^+)&
	\Bigg[\sum_{i=1}^n \left(\Tri(s_{i,n+i-2}) - \Tri(s_{i,n+i-1})\right)\nonumber\\
	-\frac{1}{2}\sum_{i=1}^n\sum_{j=i+3}^{n+i-1}\Ftme(&s_{i,j},s_{i+1,j-1};s_{i,j+1},s_{i+1,j})
	-\frac{1}{2}\sum_{i=1}^n\Fom(s_{i,i+2};s_{i,i+1},s_{i+1,i+2})\Bigg]\nonumber\\
+\frac{c_\Gamma}{\Pi_{\alpha=2}^{n}\A{\alpha}{\alpha+1}}\sum_{i=4}^n\bigg[
\frac{\NP}{6} &\langle 1P_{i,n}(i-1)2\rangle\langle 1(i-1)P_{i,2}
2\rangle\Big(\langle 1(i-1)P_{i,2}2\rangle-
\langle 1P_{i,n}(i-1)2\rangle\Big)\nonumber \\
\times & L_3(s_{i-1,1},s_{i,1})
+\frac{\beta_0}{N}\spa 1.2^2\langle 1P_{i,n}(i-1)2\rangle
L_1(s_{i-1,1},s_{i,1})\nonumber\\
+&\frac{\NP}{6} \langle 1iP_{3,i-1}2\rangle\langle 1P_{1,i-1}i2\rangle
\Big(\langle 1P_{1,i-1}i2\rangle-\langle 1iP_{3,i-1}2\rangle\Big)
\nonumber\\
\times &L_3(s_{2,i},s_{2,i-1})+\frac{\beta_0}{N}\spa 1.2^2
\langle 1iP_{3,i-1}2\rangle L_1(s_{2,i},s_{2,i-1})\bigg].
	\label{eq:1lhmmCC}
\end{align}

\subsection{Cut-completion terms}
\label{sec:completion}

The functions $L_k(s,t)$ that appear in eqs.~\eqref{eq:1lhmmCCa} 
and \eqref{eq:1lhmmCC} are bubble contributions produced 
by the reduction of tensor box and triangle integrals.
They contain unphysical singularities as $s\to t$ 
so it is useful to redefine the cut-containing
contribution in terms of a new basis which has good behaviour in the various limits. This is achieved
at the cost of adding some rational terms,
\begin{align}
	L_1(s,t) &= \wh{L}_1(s,t) \\
	L_2(s,t) &= \wh{L}_2(s,t)+\frac{1}{2(s-t)}\left(\frac{1}{t}+\frac{1}{s}\right) \\
	L_3(s,t) &= \wh{L}_3(s,t)+\frac{1}{2(s-t)^2}\left(\frac{1}{t}+\frac{1}{s}\right).
	\label{eq:newlogbasis}
\end{align}
The new $\wh{L}_k$ functions are now free from spurious singularities. Replacing the $L_k$
functions in \eqref{eq:1lhmmCC} with the corresponding $\wh{L}_k$ functions we are left with the
``completed cut term":
\begin{align}
	CR_n(\phi,1^-,&2^-,3^+,\ldots,n^+) =
\frac{c_\Gamma \NP }{12\Pi_{\alpha=2}^{n}\A{\alpha}{\alpha+1}}\nonumber\\
\sum_{i=4}^n \bigg[&\langle 1P_{i,n}(i-1)2\rangle\langle 1(i-1)P_{i,2}
2\rangle\Big(\langle 1(i-1)P_{i,2}2\rangle-
\langle 1P_{i,n}(i-1)2\rangle\Big)\nonumber \\
&\times \frac1{(s_{i-1,1}-s_{i,1})^2}\bigg(\frac1{s_{i-1,1}}
+\frac1{s_{i,1}}\bigg)\nonumber \\
+&\langle 1iP_{3,i-1}2\rangle\langle 1P_{1,i-1}i2\rangle
\Big(\langle 1P_{1,i-1}i2\rangle-\langle 1iP_{3,i-1}2\rangle\Big)\nonumber \\
&\times \frac1{(s_{2,i}-s_{2,i-1})^2}\bigg(\frac1{s_{2,i}}
+\frac1{s_{2,i-1}}\bigg)\bigg].
	\label{eq:1lhmmCR}
\end{align}

\section{The rational contributions}
\label{sec:rat}

By definition, the rational terms only contain poles in the invariants and therefore are amenable
to the same type of analysis as used for tree-amplitudes.  
We will therefore calculate the remaining rational terms in 
$A^{(1)}_n(\phi,1^-,2^-,\ldots,n^+)$ using on-shell
recursion relations that have been successful in both QCD
\cite{BDK:1lonshell,BDK:1lrecfin,Bern:bootstrap,Forde:MHVqcd,Berger:genhels,Berger:allmhv} 
and for the finite amplitudes involving the
$\phi$ field \cite{Berger:higgsrecfinite}.

This approach relies on both
the factorisation properties of one-loop amplitudes on physical poles 
and the introduction of a complex shift parameter $z$ 
to study the behaviour of the amplitude in the complex plane.

\begin{figure}[t]
	\psfrag{a}{\small$(a)$}
	\psfrag{b}{\small$(b)$}
	\psfrag{c}{\small$(c)$}
	\psfrag{f}{\small${\cal F}$}
	\begin{center}
		\includegraphics[width=1.05\textwidth]{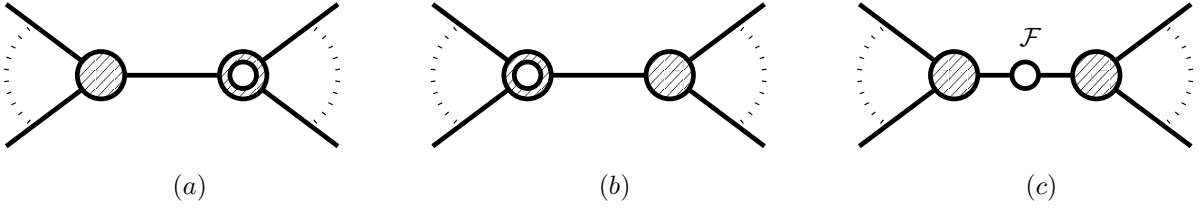}
	\end{center}
	\caption{Factorisation of one-loop amplitudes in the limit $P_i^2 \to 0$.   
	Contribution (c) does not appear for MHV amplitudes.}
	\label{fig:multipole}
\end{figure}

We recall the multi-particle factorisation properties of one-loop amplitudes \cite{Bern:1lmultifact},
\begin{equation}
\label{eq:multipole}
  A^{(1)}_n\rightarrow \frac{A_L^{(0)}A_R^{(1)}}{P_i^2}+\frac{A_L^{(1)}A_R^{(0)}}{P_i^2}
+\mathcal F\frac{A_L^{(0)}A_R^{(0)}}{P_i^2}\quad\mathrm{as}\quad P_i^2
\rightarrow 0, 
\end{equation}
where the subscripts $L$ and $R$ denote the amplitudes with fewer
external particles on the left and right of the factorising physical channel
as shown in fig.~\ref{fig:multipole}.  Note that ${\cal F}$
only contributes in multi-particle channels if the tree amplitude contains
a pole in that channel. MHV amplitudes do not have multi-particle poles and 
hence this term is absent.  

Following the discussion in the earlier sections,
we divide the amplitude into its cut-constructible and rational terms
\begin{equation}
  A^{(1)}_n =C_n +R_n.
\end{equation}
Applying eq.~\eqref{eq:multipole} in the rational sector we find (for MHV amplitudes),
\begin{eqnarray}
\label{eq:multipoleR}
R_n&\rightarrow&\frac{A_L^{(0)}R_R}{P_i^2}+\frac{R_L A_R^{(0)}}{P_i^2}
\quad\mathrm{as}\quad P_i^2
\rightarrow 0.
\end{eqnarray}

However, as it stands, $R_n$ contains unphysical poles which cancel
with unphysical poles in $C_n$.
Therefore, we add the rational function $CR_n$ (the
cut completion term of section~\ref{sec:completion}) 
to $C_n$ and subtract it from $R_n$.  $CR_n$  is chosen such that it
cancels the unphysical poles in $C_n$ (and thus, simultaneously in
$R_n$). This means that we should try to set up a recursion on 
the physical poles with
the expression $\wh{R}_n = R_n-CR_n$,

To develop the recursion, a convenient analytic continuation is to shift 
the spinors of the negative helicity gluons 1 and 2 such that
\begin{equation}
  |\widehat 1\rangle = |1\rangle +z|2\rangle,\qquad\qquad
|\widehat 2] = |2]-z|1].
\label{eq:shift12}
\end{equation}
The corresponding momenta are also shifted,
\begin{equation}
\label{eq:pshift}
p_1^{\mu} \to p_1^{\mu}(z) = p_1^{\mu} + \frac{z}{2}\langle 2 | \gamma^\mu | 1 ],
\qquad\qquad
p_2^{\mu} \to p_2^{\mu}(z) = p_2^{\mu} - \frac{z}{2}\langle 2 | \gamma^\mu | 1 ].
\end{equation} 

We now consider the integral,
\begin{equation}
\frac{1}{2\pi i} \oint_C \frac{dz}{z} \wh{R}_n(z)=
\frac{1}{2\pi i} \oint_C \frac{dz}{z} \left({R}_n(z)-CR_n(z)\right).
\end{equation}
Assuming that there is no surface term at infinity, the integral vanishes.
The remaining residues are fixed by the multiparticle factorisation 
\eqref{eq:multipole} so that the rational contribution is given by
\begin{eqnarray}
  \wh{R}_n(0)&=&-\sum_{\mathrm{phys.~poles~}z_i}{\rm Res}_{z=z_i}
  \frac{
(R_n(z)-CR_n(z))}z\nonumber\\
&=&\sum_i\frac{A_L^{(0)}(z)R_R(z)+R_L(z)A_R^{(0)}(z)}
{P_i^2}+\sum_i {\rm Res}_{z=z_i} \frac{CR_n(z)}z
\label{eq:1lrecdef}
\end{eqnarray}
The last term is called the overlap term. It is quite simple to
calculate if the poles at physical $z_i$ are all first order, as this
makes them similar to recursive terms.

To sum up, the rational terms consist of the recursive terms which are
similar in calculation to tree-level, completion terms which can be
computed simply from the cut-containing functions, and the overlap
terms which can be computed by considering the completion terms as
certain internal momenta go on-shell.

\subsection{Recursive terms \label{sec:recurs}}

\begin{figure}[t]
	\begin{center}
		\psfrag{1}{$\wh{1}^-$}
		\psfrag{2}{$\wh{2}^-$}
		\psfrag{3}{$3^+$}
		\psfrag{4}{$4^+$}
		\psfrag{ip1}{$(i+1)^+$}
		\psfrag{i-1}{$(i-1)^+$}
		\psfrag{n-1}{$(n-1)^+$}
		\psfrag{n}{$n^+$}
		\psfrag{i}{$i^+$}
		\psfrag{p}{\small$+$}
		\psfrag{m}{\small$-$}
		\psfrag{phi}{$\phi$}
		\includegraphics[width=12cm]{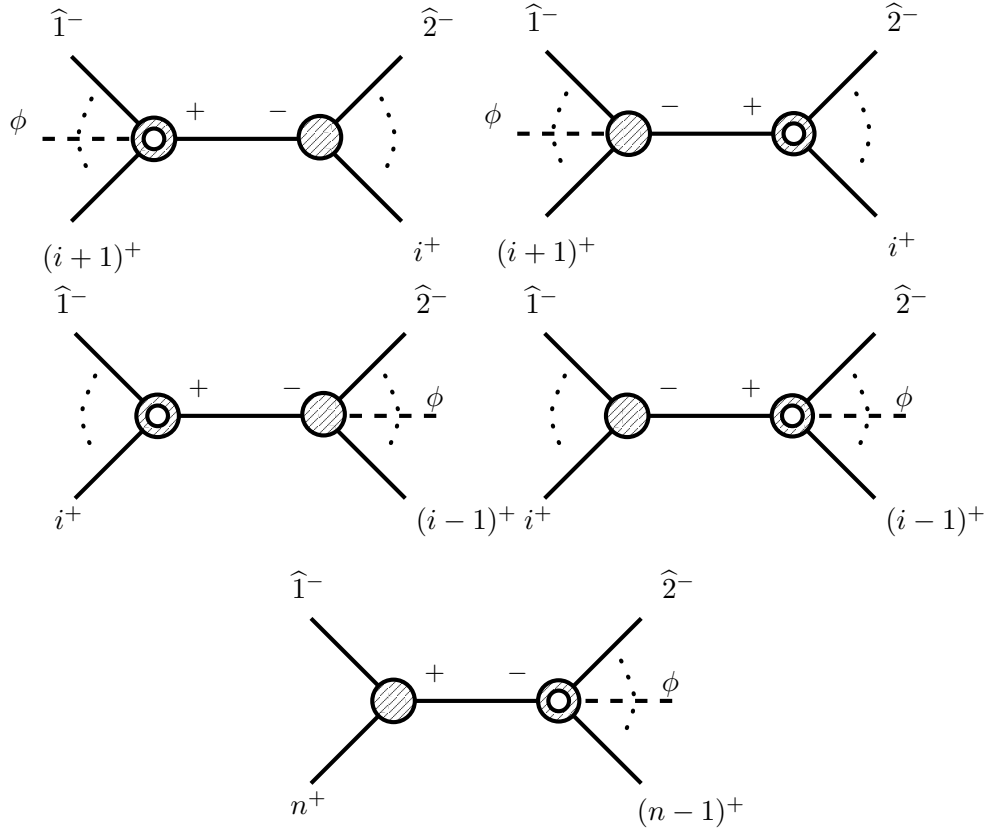}
	\end{center}
	\caption{The direct recursive diagrams contributing to $R_n(\phi,1^-,2^-,\ldots,n^+)$ with a
	$|1|\la2|$ shift. }
	\label{fig:1lrec}
\end{figure}

The recursive part of the rational contribution is defined by,
\begin{equation}
R_n^D = \sum_i\frac{A_L^{(0)}(z)R_R(z)+R_L(z)A_R^{(0)}(z)}
{P_i^2}.
\end{equation}
For the choice of shift given in eq.~(\ref{eq:shift12}) the allowed 
types of contributing diagrams contributions are shown in
fig.~\ref{fig:1lrec}. 

Combining the various diagrams, we find that recursive terms obey the following relation,
\begin{align}
	R_n^D&(\phi;1^-,2^-,3^+,\ldots,n^+) = \nonumber\\
	&\sum_{i=4}^{n-1} 
	 {R}(\phi;\wh{1}^-,\wh{P}_{2,i}^+, (i+1)^+,\ldots,n^+) 
	 \frac{1}{s_{2,i}}
	A^{(0)}(-\wh{P}_{2,i}^-,\wh{2}^-,\ldots,i^+)\nonumber\\
	+&\sum_{i=4}^{n} 
	A^{(0)}(\phi;\wh{1}^-,\wh{P}_{2,i}^-, (i+1)^+, \ldots, n^+) 
	\frac{1}{s_{2,i}}
	 {R}(-\wh{P}_{2,i}^+,\wh{2}^-,\ldots,i^+)
	 \nonumber\\
	+&\sum_{i=3}^{n-1} 
	 {R}(\wh{1}^-,-\wh{P}_{i,1}^+,i^+,\ldots,n^+) \frac{1}{s_{i,1}}
	A^{(0)}(\phi;\wh{P}_{i,1}^-,\wh{2}^-,\ldots,(i-1)^+)\nonumber\\
	+&\sum_{i=4}^{n-1} 
	A^{(0)}(\wh{1}^-,-\wh{P}_{i,1}^-,i^+,\ldots,n^+) \frac{1}{s_{i,1}}
	 {R}(\phi;\wh{P}_{i,1}^+,\wh{2}^-,\ldots,(i-1)^+)
	 \nonumber\\
	+&
	 {R}(\phi;\wh{1}^-,\wh{P}_{23}^+,4^+,\ldots,n^+) \frac{1}{s_{23}}
	A^{(0)}(-\wh{P}_{23}^-,\wh{2}^-,3^+)\nonumber\\
	+& A^{(0)}(\wh{1}^-,-\wh{P}_{n1}^+,n^+) \frac{1}{s_{n1}}
	 {R}(\phi;\wh{P}_{n1}^-,\wh{2}^-\ldots,(n-1)^+).
	\label{eq:1lrec}
\end{align}
Here ${R}$ represents the full rational part
of the one-loop amplitude with fewer external legs.  
The contributions with $i=n$ in the first term and $i=3$ in the fourth
term are absent because $A_2(\phi,-,+)$ (and hence $R_2(\phi,-,+)$) 
is zero by conservation of angular momentum.

Note that one could have written down other recursive terms 
that contain three-point pure gauge amplitudes. 
The most problematic of these are those where the
``one-loop''-ness is in the gauge three-point, since the
factorisation properties for complex momenta are not fully understood.
 However,  when the two
external gluons have opposite helicity it is guaranteed to vanish, because the
corresponding splitting function does not have rational parts. 
For our particular choice of shift, the only possible pairings of the external gluons are
($2^-$,$3^+$) or ($n^+$, $1^-$), and we find
this significant
simplification in all possible cases.

This leaves the cases where pure gauge three-point is at tree
level. Because of the way we have chosen our shift, the  amplitudes
\begin{equation}
  A^0(n^+,\widehat 1^-,-\widehat P_{n,1}^-),\qquad
A^0(\widehat 2^-,3^+,-\widehat P_{2,3}^+)
\end{equation}
both vanish, so that the only two 
contributions involving a three gluon vertex are when 
$i=3$ in the first term 
and the last term of eq.~\eqref{eq:1lrec}.
This latter contribution
is the ``homogenous'' term in the recursion; it depends on the
$\phi$-MHV amplitude with one gluon fewer.
The first two $\phi$-MHV amplitudes are known,
\begin{eqnarray}
\label{eq:ncc2-mm}
{R}_2(\phi;1^-,2^-) &=&  \frac{1}{8\pi^2}A^{(0)}(\phi,1^-,2^-),\\
\label{eq:ncc3-mmp}
{R}_3(\phi;1^-,2^-,3^+) &=& \frac{1}{8\pi^2}A^{(0)}(\phi,1^-,2^-,3^+).
\end{eqnarray}

Because the tree amplitudes with fewer than two negative
helicities vanish, the remaining one-loop contributions needed are 
those with one negative helicity. These are finite one-loop amplitudes 
and are entirely rational.  The finite $\phi-+\ldots+$ amplitudes were computed 
for arbitrary numbers of positive helicity gluons in ref.~\cite{Berger:higgsrecfinite}.
As a concrete example, the three-gluon amplitude is given by,
\begin{align}
	{R}_3(\phi;1^-,2^+,3^+) &=
	\frac{\NP}{96\pi^2}\frac{\A{1}{2}\A{3}{1}\B{2}{3}}{\A{2}{3}^2}
	-\frac{1}{8\pi^2}A^{(0)}_3(\phi^\dagger;1^-,2^+,3^+).
	\label{eq:ncc3-mpp}
\end{align}
Similarly, the pure QCD $-+\ldots+$ amplitudes are given to all orders in ref.~\cite{Mahlon:oneminus,BDK:1lonshell}.
In the four gluon case, the result is,
\begin{align}
\label{eq:ncc4-mppp}
R_4(1^-,2^+,3^+,4^+) &= \frac{\NP}{96\pi^2}
\frac{\spa 2.4 \spb 2.4^3}{\spb 1.2 \spa 2.3 \spa 3.4 \spb 4.1}
\end{align}

The value that $z$ takes is obtained by requiring that the shifted momenta  
\begin{equation}
\wh{P}^\mu_{i,j} = P^\mu_{i,j} \pm \frac{z}{2}\langle 2 | \gamma^\mu | 1 ],
\end{equation}
is on-shell.  In this equation, the sign is positive when the momentum set $\{p_i,p_j\}$ includes 
$p_1$ and is negative when it includes $p_2$.
There are four distinct channels, specified by the multiple invariants 
$s_{i,1}$, $s_{2,i}$ and the double invariants $s_{n1}$ and $s_{23}$.
In each channel, we find that the value of $z$ and the hatted variables 
are given by,
\begin{align}
	s_{i,1} \text{ channels:}
	&\qquad z_{i,1} = -\frac{s_{i,1}}{ \AB{2}{P_{i,2}}{1} },\nonumber\\
	&\qquad|\widehat 1\rangle = -\frac{|P_{i,n}P_{i,2}2\rangle}{[1P_{i,1}2\rangle},
	\quad |\widehat 2] = \frac{|P_{i,2}P_{i,n}1]}{\langle 2P_{i,1}1]},
	\quad \widehat P_{i,1}=\frac{|P_{i,n}1]\langle 2P_{i,2}|}{\langle 2P_{i,1}1]},\\
	s_{2,i} \text{ channels:}
	&\qquad z_{2,i} = \frac{s_{2,i}}{\AB{2}{P_{3,i}}{1}},\nonumber\\
	&\qquad |\widehat 1\rangle = \frac{|P_{1,i}P_{3,i}2\rangle}{[1P_{2,i}2\rangle},
	\quad |\widehat 2]=-\frac{|P_{3,i}P_{1,i}1]}{\langle 2P_{2,i}1]},
	\quad \widehat P_{2,i}=\frac{|P_{1,i}1]\langle 2P_{3,i}|}{\langle 2P_{2,i}1]},\\
	s_{n1} \text{ channel:}
	&\qquad z_{n1} = -\frac{\A{1}{n}}{\A{2}{n}},\nonumber\\
	&\qquad |\widehat 1\rangle=|n\rangle \frac{\spa 1.2}{\spa n.2},
	\quad |\widehat 2]=\frac{|P_{1,2}n\rangle}{\spa 2.n},
	\quad \widehat P_{n,1}=\frac{|n\rangle\langle 2P_{n,1}|} {\spa 2.n}, \\
	s_{23} \text{ channel:}
	&\qquad z_{23} = \frac{\B{3}{2}}{\B{3}{1}},\nonumber\\
	&\qquad   |\widehat 1\rangle =\frac{|P_{1,2}3]}{\spb 1.3},
	\quad |\widehat 2]=|3]\frac{\spb 1.2}{\spb 1.3},
	\quad \widehat P_{2,3}=\frac{|P_{2,3}1][3|}{\spb 3.1}. 
	\label{eq:shiftdefs}
\end{align}

\subsection{Overlap terms}
\label{sec:overlap}

The overlap terms are defined by~\cite{Berger:genhels}
\begin{equation}
O_n = \sum_i {\rm Res}_{z=z_i} \frac{CR_n(z)}z.
\end{equation}
They can be obtained by evaluating the residue of the cut completion term $CR_n$ given
in eq.~(\ref{eq:1lhmmCR}) in each of the physical channels.
In order to make the residue calculation straightforward, 
it is convenient to use identities such as
$s_{i,j}-s_{i,j-1} = \AB{j}{P_{i,j-1}}{j}$ to
rewrite eq.~(\ref{eq:1lhmmCR}) in a way that exposes each of
the physical poles, 
\begin{align}
	CR_n(\phi,1^-&,2^-,3^+,\ldots,n^+)=\frac{c_\Gamma\NP}{12\A
	n1 \Pi_{\alpha=2}^{n-1}\A{\alpha}{\alpha+1} }\nonumber\\
	\Bigg[
	\phantom{+}&
	\frac{\langle 1|43| 2 \rangle \langle 1|P_{2,3}4|2\rangle
	\Big(\langle 1|P_{2,3}4|2\rangle-\spa 1.4 \spb 4.3\spa3.2 \Big)}
	{s_{23} \langle 4 | P_{2,3} | 4]^2}
	\nonumber \\
	+&
	\frac{\langle 1|nP_{2,n}|2\rangle\langle 1|P_{2,n}n|2\rangle
	\Big(\langle 1|P_{2,n}n|2\rangle-\langle 1|nP_{2,n}|2\rangle\Big)}
	{s_{2,n}\langle n | P_{2,n} | n]^2}
	\nonumber\\
	 +&
	\frac{\langle 1|P_{3,1}3|2\rangle\langle 1|3P_{3,1}|
	2\rangle\Big(\langle 1|3P_{3,1}|2\rangle-\langle 1|P_{3,1}3|2\rangle\Big)}
	{s_{3,1}\langle 3| P_{3,1} | 3 ]^2}
	\nonumber \\
	+&
	\frac{\langle 1|n(n-1)|2\rangle\langle 1|(n-1)P_{n,1}|
	2\rangle\Big(\langle 1|(n-1)P_{n,1}|2\rangle-\langle 1|n(n-1)|2\rangle\Big)}
	{s_{n1}\langle (n-1)| P_{n,1} | (n-1) ]^2}	
	\nonumber\\
	+\sum_{i=4}^{n-1} & \frac{1}{s_{i,1}} \Bigg (
	\frac{\langle 1|P_{i,1}i|2\rangle\langle 1|iP_{i,1}|
	2\rangle\Big(\langle 1|iP_{i,1}|2\rangle-\langle 1|P_{i,1}i|2\rangle\Big)}
	{\langle i| P_{i,1} | i ]^2}
	\nonumber \\
	  &\phantom{\frac{1}{s_{i,1}} \Bigg (}+
	\frac{\langle 1|P_{i,1}(i-1)|2\rangle\langle 1|(i-1)P_{i,1}|
	2\rangle\Big(\langle 1|(i-1)P_{i,1}|2\rangle-\langle 1|P_{i,1}(i-1)|2\rangle\Big)}
	{\langle (i-1)| P_{i,1} | (i-1) ]^2}	\Bigg)
	\nonumber\\
	+\sum_{i=4}^{n-1} & \frac{1}{s_{2,i}}\Bigg(
	\frac{\langle 1|iP_{2,i}|2\rangle\langle 1|P_{2,i}i|2\rangle
	\Big(\langle 1|P_{2,i}i|2\rangle-\langle 1|iP_{2,i}|2\rangle\Big)}
	{s_{2,i}\langle i | P_{2,i} | i]^2}
	\nonumber\\
	  &\phantom{\frac{1}{s_{2,i}}\Bigg(}+
	\frac{\langle 1|(i+1)P_{2,i}|2\rangle\langle 1|P_{2,i}(i+1)|2\rangle
	\Big(\langle 1|P_{2,i}(i+1)|2\rangle-\langle 1|(i+1)P_{2,i}|2\rangle\Big)}
	{s_{2,i}\langle (i+1) | P_{2,i} | (i+1)]^2}\Bigg)
	\Bigg].
	\nonumber\\
	\label{eq:kasperCR}
\end{align}

We observe that the cut completion term contains
only simple residues so for the $P_{i,j}$ pole,
the overlap term is given by,
\begin{equation}
O^{i,j}_n = CR_n(z_{i,j}) \frac{\wh{s_{i,j}}}{s_{i,j}}
\end{equation}
where $z_{i,j}$ is the value of $z$ that puts $\wh{P}_{i,j}$ on-shell.
The multiplicative factor removes the $\wh{s_{i,j}}$ pole in $CR_n$ and replaces it with the
correct propagator ${s}_{i,j}$. 
Note that the only terms that are affected by the momentum shifts are $|1\rangle$, $|2]$ and any
invariant including either $p_1$ or $p_2$.   The overall factor $\A n1$ must be treated carefully, but
not $\A 23$.

Let us first consider the $s_{23}$ pole.  The overlap term is given by,
\begin{align}
	O^{23}_n &= \frac{CR(z_{23})\wh{s_{23}}}{s_{23}}\nonumber\\
	&=
\frac{c_\Gamma\NP}{12} 
\frac{\langle\wh{1} 4\rangle \spb 4.3 \langle \wh{1}|\wh{P_{2,3}}4|2\rangle
	\Big(\langle \wh{1}|\wh{P_{2,3}}4|2\rangle-\langle \wh{1} 4 \rangle\spb 4.3\spa3.2 \Big)}
	{\spb 2.3 \langle n\wh{1}\rangle \langle 4 | \wh{P_{2,3}} | 4]^2 
	\,\Pi_{\alpha=2}^{n-1}\A{\alpha}{\alpha+1} }.
\end{align}
Employing the definitions of the shifted variables in the $s_{23}$ channel given in eq.~\eqref{eq:shiftdefs},
we find that
\begin{align}
	O^{23}_n		
	&= -\frac{\NP}{192\pi^2}
	\frac{s_{123}\B 34\A 42\AB{4}{P_{12}}{3}
	\big(s_{123}\A 42+\AB{4}{P_{12}}{3}\A 32\big)
	}
	{s_{23}\AB{n}{P_{12}}{3}\AB{4}{P_{23}}{1}^2\,\Pi_{\alpha=3}^{n-1}\A{\alpha}{\alpha+1}}.
	\label{eq:O23} 
\end{align}
Similarly, we find that 
\begin{eqnarray}
	O^{2n}_n		
	&=& 
	-\frac{\NP}{192\pi^2}\frac{s_{1,n}\A 2n \AB{2}{P_{3,n}}{n}(\AA{2}{P_{3,n}P_{1,n}}{n}+s_{1,n}\A 2n )}
	{s_{2,n}\AB{n}{P_{1,n}}{1}^2\,\Pi_{\alpha=2}^{n-1}\A{\alpha}{\alpha+1}},\\
	\label{eq:O2n} 
	O^{31}_n		
	&=& \frac{\NP}{192\pi^2}\frac{s_{3,n}\A 32 \AB{2}{P_{3,1}}{3}\AA{2}{P_{3,1}P_{3,n}}{3}(\AA{2}{P_{3,1}P_{3,n}}{3}+s_{3,n}\A 23)}
	{s_{3,1}\AA{n}{P_{3,n}P_{3,1}}{2}\AB{3}{P_{3,1}}{1}^2\,\Pi_{\alpha=2}^{n-1}\A{\alpha}{\alpha+1}}.
	\label{eq:O31} 
\end{eqnarray}
The overlap in the $s_{n1}$ channel is complicated by the fact 
that there is an overall factor of $1/\A n1$. We find,
\begin{align}
	O^{n1}_n &=  \frac{\NP}{192\pi^2}
	 \frac{\A 12^3}{\A{n}{1}\Pi_{\alpha=2}^{n-1}\A{\alpha}{\alpha+1}}\Bigg[
	\frac{\B n{(n-1)}\A {(n-1)}2}{\B n1\A 12}+\nonumber\\
	\sum_{i=4}^{n-1}\Bigg\{&
	\frac{\AA{n}{P_{i,n}(i-1)}{2}\AA{n}{(i-1)P_{i,2}}{2}(\AA{n}{(i-1)P_{i,2}}{2}-\AA{n}{P_{i,n}(i-1)}{2}) }	
	{\AA{n}{(i-1)P_{i,2}+P_{i,n}(i-1)}{2}^2}
	\nonumber\\&\qquad\qquad\times
	\bigg(\frac{1}{\AA{n}{P_{i-1,n}P_{i-1,2}}{2}}+\frac{1}{\AA{n}{P_{i,n}P_{i,2}}{2}}\bigg)\nonumber\\
	&+\frac{ \AA{n}{iP_{3,i-1}}{2}\AA{n}{P_{1,i-1}i}{2}(\AA{n}{P_{1,i-1}i}{2}-\AA{n}{iP_{3,i-1}}{2}) }
	{\AA{n}{iP_{3,i-1}+P_{1,i-1}i}{2}^2}
	\nonumber\\&\qquad\qquad\times
	\bigg(\frac{1}{\AA{n}{P_{1,i}P_{3,i}}{2}}+
	\frac{1}{\AA{n}{P_{1,i-1}P_{3,i-1}}{2}}\bigg)\Bigg\}\Bigg].
	\label{eq:On1}
\end{align}

Finally, when $i \leq n-1$, the overlap terms are given by
\begin{eqnarray}
	O^{i,1}_n 
	&=&\frac{\NP}{192\pi^2}\frac{s_{i,n}}
	{s_{i,1}\AA{n}{P_{i,n}P_{i,2}}{2}\,\Pi_{\alpha=2}^{n-1}\A{\alpha}{\alpha+1}}
	\Bigg[\nonumber\\
	&&\frac{\AA{2}{P_{i,2}(i-1)}{2}\AA{2}{P_{i,2}P_{i,n}}{(i-1)}
	(\AA{2}{P_{i,2}P_{i,n}}{(i-1)}+s_{i,n}\langle 2(i-1)\rangle)}
	{\AB{(i-1)}{P_{i,n}}{1}^2}\nonumber\\
	&&+\frac{\AA{2}{P_{i,2}i}{2}\AA{2}{P_{i,2}P_{i,n}}{i}
	(\AA{2}{P_{i,2}P_{i,n}}{i}+s_{i,n}\langle 2 i\rangle )}
	{\AB{i}{P_{i,n}}{1}^2}\Bigg],\\
	\label{eq:Oi1}
	O^{2,i}_n &=& 
	 \frac{\NP}{192\pi^2}
	\frac{s_{1,i}}{s_{2,i}\AA{n}{P_{1,i}P_{3,i}}{2}\Pi_{\alpha=2}^{n-1}\A{\alpha}{\alpha+1}}
	\Bigg[\nonumber\\
	&&\frac{\A 2i \AB{2}{P_{3,i}}{i} \AA{2}{P_{3,i}P_{1,i}}{i}
	\big(\AA{2}{P_{3,i}P_{1,i}}{i} + s_{1,i}\A 2i\big)
	}{\AB{i}{P_{1,i}}{1}^2}\nonumber\\
	&&+
	\frac{\langle 2(i+1)\rangle \AB{2}{P_{3,i}}{(i+1)} \AA{2}{P_{3,i}P_{1,i}}{(i+1)}
	\big(\AA{2}{P_{3,i}P_{1,i}}{(i+1)} + s_{1,i}\langle 2(i+1)\rangle\big)
	}{\AB{(i+1)}{P_{1,i}}{1}^2}
	\Bigg].\nonumber\\
	\label{eq:O2i}
\end{eqnarray}

\section{The 4-point amplitude}
\label{sec:four}

We will now compute the rational contribution explicitly for $A^{(1)}(\phi: 1^-,2^-,3^+,4^+)$. 
The recursion relation \eqref{eq:1lrec} consists of four physical channels corresponding
to poles in the invariants $s_{23}$, $s_{234}$, $s_{41}$ and $s_{341}$. 

Using the spinor shifts given in eq.~\eqref{eq:shiftdefs} together with the known 
amplitudes given in eqs.~\eqref{eq:ncc2-mm}--\eqref{eq:ncc4-mppp}
it is straightforward to evaluate the direct recursive terms,
\begin{align}
	R_4(\phi;1^-,2^-,3^+,4^+) = R_4^{234}+R_4^{23}+R_4^{341}+R_4^{41}.
\end{align}
For example, in the $s_{234}$ channel, we find,
\begin{eqnarray}
R_4^{234}&=&A^0(\phi,\widehat 1^-,\widehat P_{234}^-)\times \frac{1}{s_{234}}
\times R(\widehat 2^-,3^+,4^+,-\widehat P_{234}^+)\nonumber \\
&=&\frac{\NP c_\Gamma}{6}\frac{\langle {\widehat 1}{\widehat P_{234}}\rangle^2
\langle 3{\widehat P_{234}}\rangle [ 3{\widehat P_{234}}]^3}{s_{234}
[{\widehat 2}3]\langle 34\rangle \langle 4{\widehat P_{234}}\rangle \B{\widehat P_{234}}{
\widehat 2}}.
\end{eqnarray}
Employing the definitions of the shifts, \eqref{eq:shiftdefs}, we easily find
\begin{eqnarray}
R_4^{234}&=&
-\frac{\NP}{96\pi^2}\frac{\spb 4.3s_{1234}^2\spa 2.4^3\langle 3P_{24}1]}
{\spa 3.4^2s_{234}\langle 2P_{34}1]\langle 4P_{23}1]^2}.
\end{eqnarray}
Similarly, 
\begin{align}
	R_4^{23} =& 
	-\frac{\NP}{96\pi^2}\frac{s_{123}\B 34\B 31\AB{4}{P_{12}}{3}}{\B 23\B 12\AB{4}{P_{23}}{1}^2}
	-\frac{1}{8\pi^2}A^{(0)}(\phi^\dagger,1^-,2^-,3^+,4^+),
	\\
	R_4^{41} =&
	\frac{1}{8\pi^2}A^{(0)}(\phi,1^-,2^-,3^+,4^+),
	\\
	R_4^{341} =&
	\frac{\NP}{96\pi^2}\frac{\B 31\AB{2}{P_{13}}{4}^3}{s_{341}\A 34\B 41^2\AB{2}{P_{34}}{1}}.
\end{align}
Notice that the expressions for $R_4^{23}$ and $R_4^{41}$ are not proportional 
to $\NP$.  It is clear
that these two terms will cancel against similar terms appearing in the
$\phi^\dagger$-\MHVb\ amplitude so that the Higgs amplitude does indeed have this property.

The overlap terms are defined by,
\begin{align}
	O_4(\phi;1^-,2^-,3^+,4^+) = O_4^{234}+O_4^{23}+O_4^{341}+O_4^{41}.
\end{align}
The individual contributions can be obtained by setting $n=4$ in eqs.~\eqref{eq:O23}, 
\eqref{eq:O2n}, \eqref{eq:O31} and \eqref{eq:On1} respectively.   After some trivial simplifications,
we find,
\begin{align}
	O_4^{23} &=
	-\frac{\NP}{192\pi^2}\frac{s_{123}\B 34\A 42 
	\big( s_{123}\A 42+\AB{4}{P_{12}}{3}\A 32 \big)
	}{s_{23}\A 34\langle 4|P_{23}|1]^2},\\
	O_4^{234} &=
	-\frac{\NP}{192\pi^2}\frac{\B 34\A42 m_H^2
	\big(m_H^2\A 42-\AA{2}{P_{34}P_{123}}{4}\big)
	}{s_{234}\A 34\langle 4P_{23}1]^2},\\
	O_4^{341} &=
	-\frac{\NP}{192\pi^2}\frac{\B 34\AB{2}{P_{13}}{4}
	\big(\AB{2}{P_{13}}{4}+\A 23\B 34\big)
	}{s_{341}\A 34\B 41^2},\\
	O_4^{41} &=
	-\frac{\NP}{192\pi^2}\frac{\B 34\A12^2}{s_{41}\A 34}.
\end{align}

With a little further algebra it is possible to write the sum of recursive and overlap terms
in a form which is free of spurious
singularities,
\begin{align}
	\wh{R}_4(\phi,1^-,&2^-,3^+,4^+) = R_4(\phi;1^-,2^-,3^+,4^+)+O_4(\phi;1^-,2^-,3^+,4^+)\nonumber \\
	&=
	\frac1{8\pi^2}A^{(0)}(A;1^-,2^-,3^+,4^+)
		+\frac{\NP\B43}{96\pi^2\A34}\bigg[\nonumber\\
		&-\frac{\A23\AB 1{P_{24}}3^2}{\A34\B43\B32s_{234}}
		+\frac{\A41\AB 3{P_{12}}3}{\A34\B12\B32}-\frac
		{\A14\AB 2{P_{13}}4^2}{\A34\B43\B41s_{341}}
		+\frac{\A32\AB 4{P_{12}}4}{\A34\B12\B41}\nonumber\\
		&+\frac{\A12^2}{\A34\B43}-\frac{\A12}{\B12}
		-\frac{\A12\AB 2{P_{13}}4}{2\B41s_{341}}+\frac{\A12
		\AB1{P_{24}}3}{2\B32s_{234}}+\frac{\A12^2}{2s_{23}}
		+\frac{\A12^2}{2s_{41}}\bigg],
	\label{eq:hmm-R4}
\end{align}
where the pseudoscalar amplitude $A^{(0)}(A;1^-,2^-,3^+,4^+)$ is given by the difference of $\phi$
and $\phi^\dagger$ amplitudes.

To form the full 1-loop $\phi$-MHV amplitude we need to add this to the cut-constructible piece
given in eq.~\eqref{eq:1lhmmCC} and the completion term of eq.~\eqref{eq:kasperCR}.
\begin{align}
	A^{(1)}_4(\phi,1^-,2^-,3^+,4^+) =&\nonumber\\
	C_4(\phi,1^-,2^-,3^+,4^+)+&CR_4(\phi,1^-,2^-,3^+,4^+)+\wh{R}_4(\phi,1^-,2^-,3^+,4^+).
\end{align}

As discussed earlier, the Higgs amplitude is constructed from the sum of 
the $\phi$ and the $\phi^\dagger$ amplitudes where the 
$\phi^\dagger$ contribution is obtained using parity symmetry,
\begin{align}
	A^{(1)}_4(H,1^-,2^-,3^+,4^+) &=
	A^{(1)}_4(\phi,1^-,2^-,3^+,4^+)+A^{(1)}_4(\phi^\dagger,1^-,2^-,3^+,4^+)\nonumber\\
	&= 
	A^{(1)}_4(\phi,1^-,2^-,3^+,4^+)+\left(A^{(1)}_4(\phi^\dagger,3^-,4^-,1^+,2^+)\right)_{\A
	ij\leftrightarrow\B ji}
\end{align}

The one-loop amplitudes in this paper are computed in the four-dimensional
helicity scheme and are not renormalised.   To perform an $\overline{MS}$
renormalisation, one should subtract an $\overline{MS}$ counterterm from
$A^{(1)}_n$,
\begin{equation}
A^{(1)}_n \to A^{(1)}_n - c_\Gamma \frac{n}{2}\frac{\beta_0}{\epsilon}
A^{(0)}_n.
\end{equation}
The Wilson coefficient \eqref{eq:C} produces an additional finite contribution,
\begin{equation}
A^{(1)}_n \to A^{(1)}_n +\frac{11}{2}~A^{(0)}_n.
\end{equation}

\section{Cross Checks and Limits}
\label{sec:checks}

\subsection{Infra-red pole structure}

The infra-red poles are constrained to have a certain form
proportional to the tree level amplitude \cite{Catani:irpoles,Giele:irpoles},
\begin{equation}
	A^{(1)}_n = -\frac{c_\Gamma}{\e^2} A^{(0)}_n\sum_{i=1}^{n}
	\left(\frac{\mu^2}{-s_{ii+1}}\right)^{\e} + \mc O(\e^0).
	\label{eq:1lpoles2}
\end{equation}
Expanding the hypergeometric functions as a series in $\e$ quickly leads to a proof of this fact,
and it can be seen that all logarithms vanish at order $1/\e$.

\subsection{Collinear Limits}

In general the collinear behaviour of one-loop amplitudes can be written\cite{BDDK:uni1,Kosower:allcoll}:
\begin{align}
	A^{(1)}_n(\ldots,&i^{\lambda_i},i+1^{\lambda_{i+1}},\ldots)
	\overset{i||i+1}{\to}\nonumber\\ 
	\sum_{h=\pm}
	& 
	A^{(1)}_{n-1}(\ldots,{i-1}^{\lambda_{i-1}},P^h,i+2^{\lambda_{i+2}},\ldots)\Split^{(0)}(-P^{-h};i^{\lambda_i},i+1^{\lambda_{i+1}})\nonumber\\
	+& 
	A^{(0)}_{n-1}(\ldots,{i-1}^{\lambda_{i-1}},P^h,i+2^{\lambda_{i+2}},\ldots)\Split^{(1)}(-P^{-h};i^{\lambda_i},i+1^{\lambda_{i+1}})	
	\label{eq:2coll1l}
\end{align}
where the collinear limit is defined through $p_i \to zP$ and $p_{i+1}\to (1-z)P$. The universal
splitting functions for QCD have been calculated in reference
\cite{BDDK:uni1,BDDK:uni2,Bern:allorder}. The tree-level splitting
functions are given by,
\begin{align}
	\Split^{(0)}(-P^+,1^-,2^+) &= \frac{z^2}{\sqrt{ z(1-z)}\A{1}{2} }\\
	\Split^{(0)}(-P^+,1^+,2^-) &= \frac{(1-z)^2}{\sqrt{ z(1-z)}\A{1}{2} }\\
	\Split^{(0)}(-P^-,1^+,2^+) &= \frac{1}{\sqrt{ z(1-z)}\A{1}{2} }\\
	\Split^{(0)}(-P^-,1^-,2^-) &= 0. \label{eq:split0mmm}
\end{align}
It is convenient to divide the one-loop splitting functions into cut-constructible and
rational components,
\begin{equation}
	\Split^{(1)}(-P^{-h},1^{\lambda_1},2^{\lambda_2}) =
	\Split^{(1),C}(-P^{-h},1^{\lambda_1},2^{\lambda_2})+\Split^{(1),R}(-P^{-h},1^{\lambda_1},2^{\lambda_2})
\end{equation}
where
\begin{align}
	&\Split^{(1),C}(-P^\pm,1^-,2^+) =
	\Split^{(0)}(-P^\pm,1^-,2^+)\frac{c_\Gamma}{\e^2}\times\nonumber\\
	&\hspace{3mm}\left( \frac{\mu^2}{-s_{12} } \right)^{\e}
	\left( 1 - \FF{ 1,-\e;1-\e;
	\frac{z}{z-1} } - \FF{ 1,-\e;1-\e;\frac{z-1}{z}  } \right),\\
	&\Split^{(1),C}(-P^+,1^-,2^-) =
	\Split^{(0)}(-P^+,1^-,2^-)\frac{c_\Gamma}{\e^2}\times\nonumber\\
	&\hspace{3mm}\left( \frac{\mu^2}{-s_{12} } \right)^{\e}
	\left( 1 - \FF{ 1,-\e;1-\e;
	\frac{z}{z-1} } - \FF{ 1,-\e;1-\e;\frac{z-1}{z}  } \right),\\
	&\Split^{(1),C}(-P^-,1^-,2^-) = 0, \\ 
	&\Split^{(1),R}(-P^\pm,1^-,2^+)  =  0, \\
	&\Split^{(1),R}(-P^+,1^-,2^-)  =  \frac{\NP}{96\pi^2}\frac{\sqrt{z(1-z)}}{\B{1}{2}}, \\
	&\Split^{(1),R}(-P^-,1^-,2^-) = \frac{\NP}{96\pi^2}\frac{ \sqrt{z(1-z)}\A{1}{2} }{\B{1}{2}^2}. 
	\label{eq:split1l}
\end{align}

\subsection{Collinear factorisation of the cut-constructible contributions}

Considering only the cut constructible contributions, we expect the following limiting behaviour
\begin{align}
	C^{(1)}_n(\ldots,&i^{\lambda_i},i+1^{\lambda_{i+1}},\ldots)
	\overset{i||i+1}{\to}\nonumber\\ 
	\sum_{h=\pm}
	& 
	C^{(1)}_{n-1}(\ldots,{i-1}^{\lambda_{i-1}},P^h,i+2^{\lambda_{i+2}},\ldots)\Split^{(0)}(-P^{-h};i^{\lambda_i},i+1^{\lambda_{i+1}})\nonumber\\
	+& 
	A^{(0)}_{n-1}(\ldots,{i-1}^{\lambda_{i-1}},P^h,i+2^{\lambda_{i+2}},\ldots)\Split^{(1),C}(-P^{-h};i^{\lambda_i},i+1^{\lambda_{i+1}})	
	\label{eq:2coll1l-C}
\end{align}
There are three collinear limits to consider for the $\phi$-MHV amplitude; two collinear negative helicity
gluons, two collinear positive helicity gluons and the mixed case with one of each helicity.

We first note that the box and triangle scalar integrals always appear in the combination
\begin{align}
	&U_n = 
	\Bigg[\sum_{i=1}^n \left(F^{1m}_3(s_{i,n+i-2}) - F^{1m}_3(s_{i,n+i-1})\right)\nonumber\\
	&-\frac{1}{2}\sum_{i=1}^n\sum_{j=i+3}^{n+i-1}F^{2me}_4(s_{i,j},s_{i+1,j-1};s_{i,j+1},s_{i+1,j})
	-\frac{1}{2}\sum_{i=1}^nF^{1m}_4(s_{i,i+2};s_{i,i+1},s_{i+1,i+2})\Bigg]
	\label{eq:CCpoles}
\end{align}
and multiply the tree amplitude.  

\begin{figure}[t!]
	\begin{center}
		\psfrag{ip2}{\scriptsize$(i+2)$}
		\psfrag{ip1}{\scriptsize$(i+1)$}
		\psfrag{i}{\scriptsize$i$}
		\psfrag{i}{\scriptsize$i$}
		\psfrag{jm1}{\scriptsize$(j-1)$}
		\psfrag{j}{\scriptsize$j$}
		\psfrag{P}{\scriptsize$P$}
		\psfrag{i-1}{\scriptsize$i-1$}
		\psfrag{i+1}{\scriptsize$i+1$}
		\psfrag{i+2}{\scriptsize$i+2$}
		\psfrag{i||i+1}{\scriptsize$i|i+1$}
		\includegraphics[width=10cm]{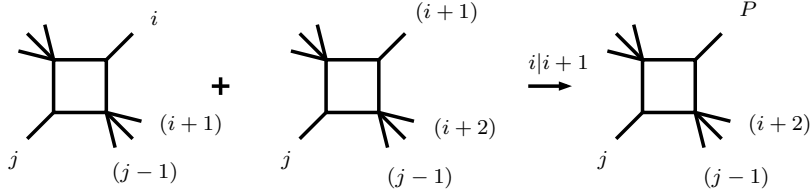}
	\end{center}
	\caption{Factorisation of combinations of two-mass ``easy" box functions 
	in limit where $i || i+1$.}
	\label{fig:boxfacta}
\end{figure}
Many of the collinear factorisation properties of the individual box and triangle functions have been
studied in Ref.~\cite{BDDK:uni1} for example,
\begin{align}
	F_4^{2me}&(s_{i,j},s_{i+1,j-1};s_{i,j-1},s_{i+1,j})+F_4^{2me}(s_{i+1,j},s_{i+2,j-1};s_{i+1,j-1},s_{i+2,j})\nonumber\\ 
	&\hspace{2cm}\overset{i||i+1}{\to} F_4^{2me}(s_{P,j},s_{i+2,j-1};s_{i+2,j},s_{P,j-1}), 
\end{align}
where $s_{P,j} = (P+p_{i+2}+\ldots+p_j)^2$ and which is illustrated in Fig.~\ref{fig:boxfacta}.
A similar relation applies when $j=i+3$ such that the second and third terms are one-mass boxes. 

\begin{figure}[t!]
	\begin{center}
		\psfrag{nm1}{\scriptsize$(n-1)$}
		\psfrag{n}{\scriptsize$n$}
		\psfrag{ip2}{\scriptsize$(i+2)$}
		\psfrag{ip1}{\scriptsize$(i+1)$}
		\psfrag{im1}{\scriptsize$(i-1)$}
		\psfrag{i}{\scriptsize$i$}
		\psfrag{i}{\scriptsize$i$}
		\psfrag{j}{\scriptsize$j$}
		\psfrag{P}{\scriptsize$P$}
		\psfrag{i-1}{\scriptsize$i-1$}
		\psfrag{i+1}{\scriptsize$i+1$}
		\psfrag{i+2}{\scriptsize$i+2$}
		\psfrag{i||i+1}{\scriptsize$i|i+1$}
		\psfrag{1}{\scriptsize$1$}
		\includegraphics[width=12cm]{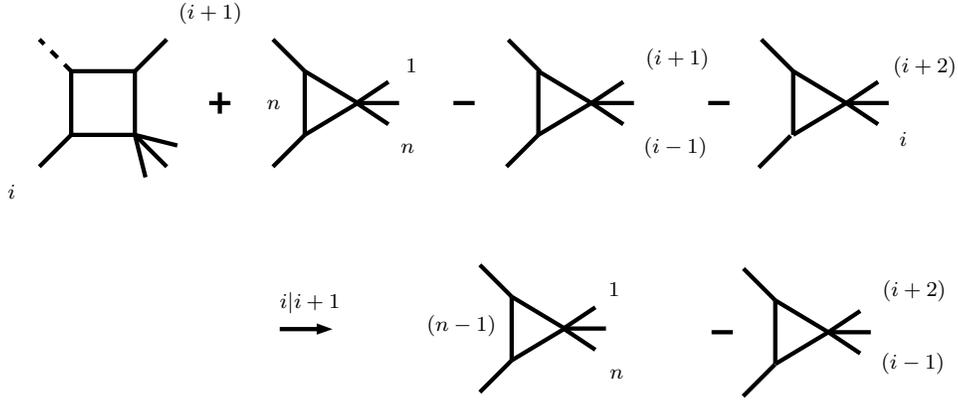}
	\end{center}
	\caption{Factorisation of a combination of two-mass ``easy" box   and one-mass triangle functions in limit where $i || i+1$.
	Note that the scale relevant for the triangle functions is indicated on the right of the triangle.  The massless legs on the left
	of the triangles do not correspond to physical momenta.}
	\label{fig:boxfactb}
\end{figure}
In much the same way, the particular combination of box and one-mass triangle functions shown in Fig.~\ref{fig:boxfactb}
factorise onto triangle functions as follows,
\begin{align}
	F^{2me}&(s_{1,n},s_{i+2,i-1};s_{i+1,i-1},s_{i+2,i}) + nF_3^{1m}(s_{1,n}) - F_3^{1m}(s_{i+1,i-1}) - F_3^{1m}(s_{i+2,i})\nonumber\\
	&\hspace{2cm}\overset{i||i+1}{\to} (n-1)F_3^{1m}(s_{1,n})-F_3^{1m}(s_{i+2,i-1}).
\end{align}

Taken together, it is straightforward to find the collinear limit of $U_n$ 
\begin{align}
\label{eq:ufact}
	U_n(&1,\ldots,i,i+1,\ldots,n) \overset{i||i+1}{\to} U_{n-1}(1,\ldots,P,\ldots,n)\nonumber\\
	&+\frac{1}{\e^2}\left(\frac{\mu^2}{-s_{i,i+1}}\right)^\e
	\left( 1-\FF{1,-\e;1-\e;\frac{z}{z-1}}-\FF{1,-\e;1-\e;\frac{z-1}{z}} \right),
\end{align}
which is independent of which (adjacent) pair are taken collinear.

The remaining limits depend on the helicity of the collinear gluons.
To explore the limits in detail, we find it convenient to utilise 
the expression for $C_n$ given in
eq.~\eqref{eq:1lhmmCCa}.

\subsubsection{The mixed helicity collinear limit}

\begin{figure}
	\psfrag{phi}{\small$\phi$}
	\psfrag{m}{\tiny$-$}
	\psfrag{p}{\tiny$+$}
	\psfrag{1}{\small$1^-$}
	\psfrag{2}{\small$2^-$}
	\psfrag{3}{\small$3^+$}
	\psfrag{4}{\small$4^+$}
	\psfrag{n}{\small$n^+$}
	\psfrag{nm1}{\small$(n-1)^+$}
	\psfrag{2||3}{\hspace{-1mm}\small$2||3$}
	\begin{center}
		\includegraphics[width=\textwidth]{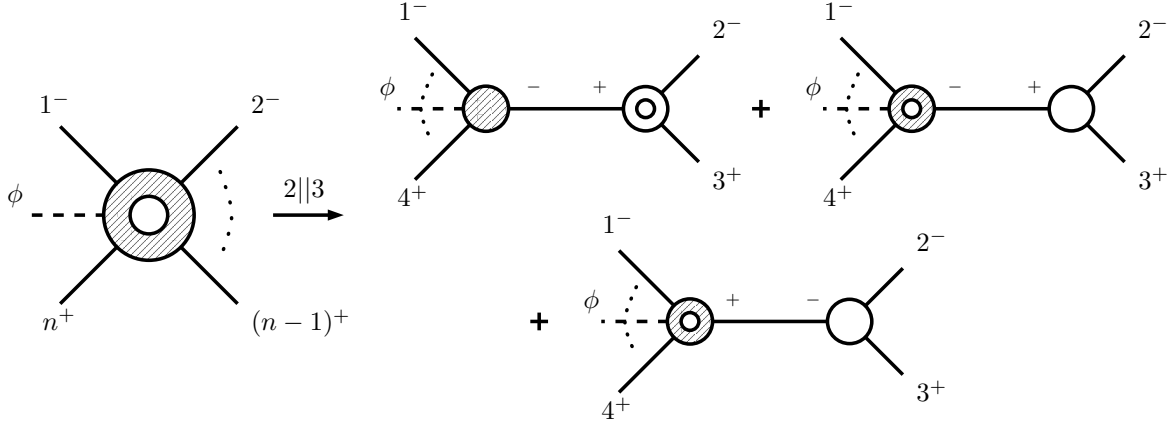}
	\end{center}
	\caption[Collinear factorisation of $A^{(1)}(\phi;1^-,2^-,3^+,\ldots,n^+)$ taking $p_2$ and
	$p_3$ parallel]
	{Collinear factorisation of $A^{(1)}(\phi;1^-,2^-,3^+,\ldots,n^+)$ taking $p_2$ and
	$p_3$ parallel.}
	\label{fig:1lhcoll2}
\end{figure}

Let us now consider
the case where we take a negative and a positive helicity collinear, e.g. the $2||3$ limit shown in
fig.~\ref{fig:1lhcoll2}.
We immediately see that the third diagram involves a factorisation onto a 
finite $\phi$ amplitude which has no cut-constructible
part and therefore gives no contribution.

The finite logarithms 
also factorise in the expected way
since,
\begin{align}
	\frac{\trm(1P_{i,n}(i-1)2)}{s_{12}} &\overset{2||3}{\to} \frac{\trm(1P_{i,n}(i-1)P)}{s_{1P}}
	& \, i>4, \\
	\frac{\trm(2P_{3,i-1}i1)}{s_{12}} &\overset{2||3}{\to} \frac{\trm(PP_{4,i-1}i1)}{s_{1P}}
	& \, i>4,
\end{align}
while
\begin{align}
	&\frac{\trm(1P_{4,n}32))}{s_{12}} \overset{2||3}{\to} 0 \\
	&\frac{\trm(2341))}{s_{12}} \overset{2||3}{\to} 0
\end{align}
ensures that terms with divergent logarithms, i.e. $L_k(s_{234},s_{23})$, will always be proportional to a trace
which vanishes in the limit and hence do not appear in the one-loop splitting function. 
Together with eq.~\eqref{eq:ufact}, we find that, as expected, 
\begin{align}
	C_n(\phi;1^-,2^-,3^+,\ldots,n^+)& \overset{2||3}{\to}\nonumber\\ 
	&A_{n-1}^{(0)}(\phi;1^-,P^-,4^+,\ldots,n^+)\Split^{(1),C}(-P^+,2^-,3^+)\nonumber\\
	+&C_{n-1}(\phi;1^-,P^-,4^+,\ldots,n^+) \Split^{(0)}(-P^+,2^-,3^+).
	\label{eq:23limit}
\end{align}

\subsubsection{Two collinear negative helicity gluons}

\begin{figure}
	\psfrag{phi}{\small$\phi$}
	\psfrag{m}{\tiny$-$}
	\psfrag{p}{\tiny$+$}
	\psfrag{1}{\small$1^-$}
	\psfrag{2}{\small$2^-$}
	\psfrag{3}{\small$3^+$}
	\psfrag{4}{\small$4^+$}
	\psfrag{n}{\small$n^+$}
	\psfrag{nm1}{\small$(n-1)^+$}
	\psfrag{1||2}{\hspace{-1mm}\small$1||2$}
	\begin{center}
		\includegraphics[width=0.6\textwidth]{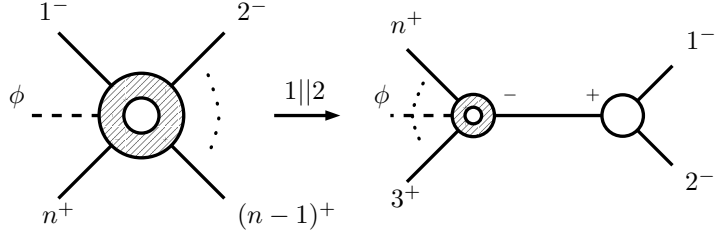}
	\end{center}
	\caption[Collinear factorisation of $A^{(1)}(\phi;1^-,2^-,3^+,\ldots,n^+)$ taking $p_1$ and
	$p_2$ parallel]
	{Collinear factorisation of $A^{(1)}(\phi;1^-,2^-,3^+,\ldots,n^+)$ taking $p_1$ and
	$p_2$ parallel. The only contribution is finite so the cut-constructible
	parts vanish in this limit.}
	\label{fig:1lhcoll3}
\end{figure}

The $1||2$ limit is rather trivial since there is only a single, finite term in this case as shown
in fig.~\ref{fig:1lhcoll3}. The tree amplitude in this case vanishes,
\begin{equation}
	A^{(0)}_n(1^-,2^-,3^+,\ldots,n^+) \overset{1||2}{\to} 0,
\end{equation}
therefore it is obvious that all the box and triangle terms of eq.~\eqref{eq:1lhmmCC} will vanish.
The remaining finite logs appear to have a singularity in $s_{12}$, the worst coming from the traces
raised to the 3rd power,
\begin{equation}
	\frac{\trm(1XY2)^3}{s_{12}^3} = \frac{\AA{1}{XY}{2}^3}{\A{1}{2}^3}\qquad,\qquad
	\frac{\trm(2XY1)^3}{s_{12}^3} = -\frac{\AA{2}{XY}{1}^3}{\A{1}{2}^3}.
\end{equation}
However the tree amplitude is proportional to $\A{1}{2}^3$ so
$C_n(\phi,1^-,2^-,3^+,\ldots,n^+)$ vanishes in the limit as expected.

\subsubsection{Two collinear positive helicity gluons}

\begin{figure}
	\psfrag{phi}{\small$\phi$}
	\psfrag{m}{\tiny$-$}
	\psfrag{p}{\tiny$+$}
	\psfrag{1}{\small$1^-$}
	\psfrag{2}{\small$2^-$}
	\psfrag{3}{\small$3^+$}
	\psfrag{4}{\small$4^+$}
	\psfrag{n}{\small$n^+$}
	\psfrag{nm1}{\small$(n-1)^+$}
	\psfrag{p||p}{\hspace{-1mm}\small$+||+$}
	\begin{center}
		\includegraphics[width=\textwidth]{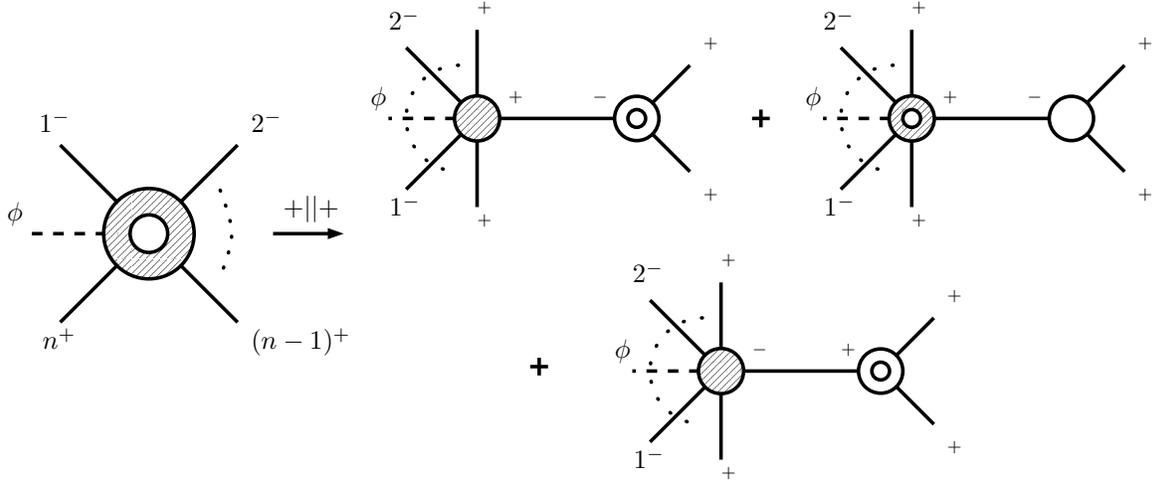}
	\end{center}
	\caption[Collinear factorisation of $A^{(1)}(\phi;1^-,2^-,3^+,\ldots,n^+)$ taking any two
	positive helicities parallel]
	{Collinear factorisation of $A^{(1)}(\phi;1^-,2^-,3^+,\ldots,n^+)$ taking any two
	positive helicities parallel.}
	\label{fig:1lhcoll4}
\end{figure}

The final limit occurs when we take any adjacent pair of
positive helicities collinear. The relevant diagrams are shown in fig.~\ref{fig:1lhcoll4}. 
We can drop the third contribution since it 
involves the purely rational splitting function.
 
We must also be able to show that there are no divergent terms coming from the finite
logs and that these terms correctly factorise onto the lower point amplitude. This turns out to be
slightly more involved than in the $2||3$ case. First let us choose to take two adjacent particles
$a$ and $b$ collinear where $p_a$ lies to the left of $p_b$ in the clockwise ordering. Using,
\begin{align}
	s_{b,i} &\overset{a||b}{\to} (1-z)s_{P,i}+zs_{b+1,i}, \\
	s_{i,a} &\overset{a||b}{\to} zs_{i,P}+(1-z)s_{i,a-1},
\end{align}
it is then possible to show:
\begin{align}
	\frac{\trm(2P_{3,a-1}a1)^k}{s_{12}^k} L_k(s_{2,a},s_{2,a-1})
	&+\frac{\trm(2P_{3,a}b1)^k}{s_{12}^k} L_k(s_{2,b},s_{2,a})\nonumber\\
	&\overset{a||b}{\to}
	\frac{\trm(2P_{3,a-1}P1)^k}{s_{12}^k} L_k(s_{2,P},s_{2,a-1}),
\end{align}
and,
\begin{align}
	\frac{\trm(1P_{b+1,i}b2)^k}{s_{12}^k} L_k(s_{b,i},s_{b+1,i})
	&+\frac{\trm(1P_{b,i}a2)^k}{s_{12}^k} L_k(s_{a,i},s_{b,i})\nonumber\\
	&\overset{a||b}{\to}
	\frac{\trm(1P_{b+1,i}P2)^k}{s_{12}^k} L_k(s_{P,i},s_{b+1,i}).
\end{align}
Note that $s_{2,P} = (p_2+\ldots+p_{a-1}+P)^2$ and $s_{P,i}=(P+p_{b+1}+\ldots+p_i)^2$.
Using these identities and recognising that $\trm(1ab2)\overset{a||b}{\to} 0$ and together
with eq.~\eqref{eq:ufact}, it is straightforward to show
that eq.~\eqref{eq:1lhmmCC} has the correct factorisation properties,
\begin{align}
	&C_n(\phi;1^-,2^-,3^+,\ldots,a^+,b^+,\ldots,n^+) \overset{a||b}{\to}\nonumber\\ 
	&C_{n-1}(\phi;1^-,2^-,3^+,\ldots,a-1^+,P^+,b+1^+,\ldots,n^+)
	\Split^{(0)}(-P^-,a^+,b^+)\nonumber\\
	+&\mc
	A_{n-1}^{(0)}(\phi;1^-,2^-,3^+,\ldots,a-1^+,P^+,b+1^+,\ldots,n^+)\Split^{(1),C}(-P^-,a^+,b^+).
	\label{eq:++limit}
\end{align}

\subsubsection{Collinear factorisation of the rational contributions}

In this section, we focus on the collinear limit of the 
rational part of the 4 gluon amplitude with two negative and two positive
helicities.   There are three independent collinear limits, $1||2$, $2||3$ and $3||4$. 

Let us first consider the case where $1^-$ and $2^-$ are parallel. 
From fig.~\ref{fig:1lhcoll3} we see that there is only a single contributing term and 
it is fairly straightforward to show that eq.~\eqref{eq:hmm-R4} satisfies,
\begin{align}
	&R_4(\phi;1^-,2^-,3^+,4^+) \overset{1||2}{\to}  
	 \frac{1}{8\pi^2} {R}_3(\phi^\dagger;P^-,3^+,4^+) \Split^{(0)}(-P^+,1^-,2^-).
	\label{eq:12ncclim}
\end{align}

The $2^-||3^+$ limit, shown in fig.~\ref{fig:1lhcoll2}, has only two contributing diagrams for the
raitonal piece since the first contribution vanishes since the one-loop mixed helicity
splitting function has no rational part.  The rational part of the four gluon amplitude
can easily be shown to satisfy,
\begin{align}
	&{R}_4(\phi;1^-,2^-,3^+,4^+)+CR_4(\phi;1^-,2^-,3^+,4^+) \overset{2||3}{\to}\nonumber\\ 
	+&{R}_3(\phi^\dagger;1^-,P^+,4^+) \Split^{(0)}(-P^-,2^-,3^+)\nonumber\\
	+&{R}_3(\phi;1^-,P^-,4^+) \Split^{(0)}(-P^+,2^-,3^+)\nonumber\\
	\label{eq:23ncclim}
\end{align}

The final collinear limit is given by taking $3^+$ and $4^+$ collinear. Here we again find the
expected behaviour,
\begin{align}
	&R_4(\phi;1^-,2^-,3^+,4^+) \overset{3||4}{\to}\nonumber\\ 
	&R_3(\phi;1^-,2^-,P^+) \Split^{(0)}(-P^-,3^+,4^+)\nonumber\\
	+&A^{(0)}(\phi;1^-,2^-,P^+)\Split^{(1),R}(-P^-,3^+,4^+)\nonumber\\
	+&A^{(0)}(\phi;1^-,2^-,P^-)\Split^{(1),R}(-P^+,3^+,4^+).
	\label{eq:34ncclim}
\end{align}

\subsection{Soft Higgs Limit}

For the case of a massless Higgs boson, we can consider the kinematic limit $p_H \to 0$.
Because of the form of the $H G_{\mu\nu}G^{\mu\nu}$ interaction, the Higgs field
behaves like a constant in this limit, so the Higgs-plus-$n$-gluon amplitudes should be related to
pure gauge theory amplitudes. Low energy theorems relate the
amplitudes with zero Higgs momentum to pure gauge theory amplitudes~\cite{Gunion:higgshunters}:
\begin{equation}
	A^{(l)}_n(H,\{g_i,\lambda_i\}) \overset{p_H\to 0}{\to} Cg\frac{\partial}{\partial g}
	A^{(l)}_n(\{g_i,\lambda_i\}),
	\label{eq:lowenergy}
\end{equation}
where $C$ is the effective coupling of Higgs field to the gluon fields.
The $n$-gluon tree amplitude is proportional to $g^{n-2}$ (see equation
\eqref{TreeColorDecomposition}) therefore,
\begin{equation}
	A^{(0)}_n(H,\{g_i,\lambda_i\}) \overset{p_H\to 0}{\to} \text{(const.)}\times(n-2)
	A^{(0)}_n(\{g_i,\lambda_i\}).	
\end{equation}
In terms of the $\phi$ and $\phi^\dagger$ components~\cite{Dixon:MHVhiggs},
\begin{align}
	&A^{(0)}_n(\phi,\{g_i,\lambda_i\}) \overset{p_\phi\to 0}{\to} \text{(const.)}\times (n_--1)
	A^{(0)}_n(\{g_i,\lambda_i\}),\\
	&A^{(0)}_n(\phi^\dagger,\{g_i,\lambda_i\}) \overset{p_{\phi^\dagger}\to 0}{\to}
	\text{(const.)}\times (n_+-1)
	A^{(0)}_n(\{g_i,\lambda_i\})	
\end{align}
where $n_+$ and $n_-$ are the number of positive and negative helicity particles respectively.

The one-loop amplitudes are proportional to $g^n$ hence similarly one can deduce the following
behaviour in the soft Higgs limit,
\begin{equation}
	A^{(1)}_n(H,\{g_i,\lambda_i\}) \overset{p_H\to 0}{\to} \text{(const.)}\times n
	A^{(1)}_n(\{g_i,\lambda_i\}).	
\end{equation}
It has been conjectured in~\cite{Berger:higgsrecfinite} that the $\phi$ and $\phi^\dagger$
components generalise from the tree level relations to give,
\begin{align}
	&A^{(1)}_n(\phi,\{g_i,\lambda_i\}) \overset{p_\phi\to 0}{\to} \text{(const.)}\times n_-
	A^{(1)}_n(\{g_i,\lambda_i\}),\\
	&A^{(1)}_n(\phi^\dagger,\{g_i,\lambda_i\}) \overset{p_{\phi^\dagger}\to 0}{\to}
	\text{(const.)}\times n_+
	A^{(1)}_n(\{g_i,\lambda_i\}).	
\end{align}

The 4-gluon MHV amplitude at one-loop in QCD has been derived in ref.~\cite{Bern:1l4g} and is given,
unrenormalised in the four-dimensional helicity scheme by
\begin{align}
	 C_4(1^-,2^-,3^+,4^+)  = &
	 -\frac{2c_\Gamma}{\e^2}A^{(0)}(1^-,2^-,3^+,4^+)\Bigg[
	\left( \frac{\mu^2}{-s_{14}} \right)^\e
	+\left( \frac{\mu^2}{-s_{12}} \right)^\e
	+ \log^2\left( \frac{s_{12}}{s_{14}} \right) + \pi^2 \nonumber\\
	- &\frac{\beta_0}{N\e(1-2\e)}\left( \frac{\mu^2}{-s_{14}} \right)^\e\Bigg],\\
	 R_4(1^-,2^-,3^+,4^+)  = & \frac{\NP c_\Gamma}{9}A^{(0)}(1^-,2^-,3^+,4^+).
\end{align}

\subsubsection{Soft limit of $A^{(1)}_4(\phi,1^-,2^-,3^+,4^+)$}

Firstly let us consider the soft limit of the cut constructible components, eq.~\eqref{eq:1lhmmCC}.
The 1-mass and 2-mass easy box functions and triangle functions have
smooth soft limits where we take:
\begin{align}
	\left( \frac{\mu^2}{-m_\phi^2} \right)^\e \overset{p_\phi\to 0}{\to} 0 \\
	\left( \frac{\mu^2}{-s_{\phi i}} \right)^\e \overset{p_\phi\to 0}{\to} 0.
\end{align}
We must apply the same relations to the finite logs coming from the tensor triangle integrals, for
instance we find:
\begin{align}
	L_k(s_{341},s_{41}) = \frac{ {\rm Bub}(s_{341}) - {\rm Bub}(s_{41})}{(s_{341}-s_{41})^k}
	\overset{p_\phi\to 0}{\longrightarrow}
	\frac{(-1)^k}{s_{14}^k\e(1-2\e)}\left( \frac{\mu^2}{-s_{14}} \right)^\e.
\end{align}
Applying these relations together with momentum conservation to the cut-constructible part of the 4
gluon amplitude we find that the boxes and
triangle functions collapse onto the correct $1/\e^2$ poles while the coefficients of
the bubble contributions simplify considerably, e.g.,
\begin{equation}
	\left(\frac{\trm(1432)}{s_{12}}\right)^k \overset{p_\phi\to 0}{\longrightarrow} (-1)^k s_{14}^k,
\end{equation}
so that all of the logarithms proportional to $\NP$ cancel among themselves leaving, 
\begin{equation}
	\left( \frac{\mu^2}{-s_{14}} \right)^\e
	\frac{2\beta_0}{N\e(1-2\e)}.
\end{equation}
Combining this with the $1/\e^2$ poles we find,
\begin{align}
	 C_4(\phi,1^-,2^-,3^+,4^+) &\overset{p_\phi\to 0}{\longrightarrow} 2 ~C_4(1^-,2^-,3^+,4^+)
\end{align}
The soft limit of the rational part, eq.~\eqref{eq:hmm-R4} and eq.~\eqref{eq:1lhmmCR}, gives
\begin{equation}
	R_4(\phi;1^-,2^-,3^+,4^+) \overset{p_\phi\to 0}{\longrightarrow} -\frac{N_p c_\Gamma}{3}A^{(0)}(1^-,2^-,3^+,4^+),
\end{equation}
with all the poles in the triple invariants vanishing. This implies that the rational terms vanish in the limit
for the full Higgs field,
\begin{equation}
	R_4(H;1^-,2^-,3^+,4^+) \overset{p_\phi\to 0}{\longrightarrow} 0,
\end{equation}
since the $\phi^\dagger$ contribution appears with the opposite sign. This confirms that
cut-constructible terms of the amplitude do follow the naive factorisation proposed in
\cite{Berger:higgsrecfinite} but the rational parts do not. This seems to be consistent with the
results for the $Hq\bar{q}Q\bar{Q}$ amplitude presented in \cite{Ellis:1lh24}.

\section{Conclusions}
\label{sec:conclusions}

Recent developments based on unitarity and on-shell recursion relations
have lead to important breakthroughs in computing non-supersymmetric 
one-loop gauge theory amplitudes. In particular, new compact analytic results
have been obtained for gluonic amplitudes involving six or more gluons.

In this paper we have focused on amplitudes involving gluons and 
a colourless scalar - the Higgs boson.
The model which we use to calculate these amplitudes is
the tree-level pure gauge theory plus an effective interaction
$HG_{\mu\nu}G^{\mu\nu}$ produced by considering the 
heavy top quark limit of the one-loop coupling of
Higgs and gluons in the non-supersymmetric standard model.
 Following Ref.~\cite{Dixon:MHVhiggs}, we split the interaction into 
selfdual and anti-selfdual pieces.  The self-dual (anti-self-dual)
gauge fields interact with the 
$\phi$ ($\phi^\dagger$) scalars respectively and  because of this selfduality, 
the amplitudes for $\phi$ plus $n$ gluons, and those for $\phi^\dagger$ plus $n$ gluons,
each have a simpler structure than the gluonic amplitudes for
either $H$ or $A$.

Previous studies using the unitarity-factorisation booststrap program
have focused on the $\phi$ amplitudes that are finite at one-loop, i.e. 
the all positive or nearly all positive helicity~\cite{Berger:higgsrecfinite}
or the divergent amplitudes when the gluons all have negative helicity~\cite{Badger:1lhiggsallm}. 
In this paper we have employed four-dimensional unitarity and 
recursion relations to compute the one-loop corrections to amplitudes
involving a colourless scalar $\phi$, two colour adjacent negative helicity gluons and 
an arbitrary number of positive helicity gluons - the so-called $\phi$-MHV amplitudes.
 
The gluonic production of Higgs bosons via a heavy quark loop is expected to be the largest
source of Higgs bosons at the LHC.    Because of the size of the strong coupling, it will be
important to understand Higgs plus jet events in some detail. The two jet channel has contributions
of both QCD~\cite{Dawson:Htomultijet,Kauffman:H2jets,DelDuca:Hto3jets,
DelDuca:H2j1,DelDuca:H2j2,DelDuca:H2j3,Klamke:LHChjj,Campbell:NLOHjj} and weak boson fusion
origin~\cite{FIGY:nloWBF1,FIGY:nloWBF2,Berger:WBF}. Separating these two ``signals" is of crucial
importance for measuring the coupling of the Higgs bosons to standard model particles from
LHC data~\cite{Duhrssen:Hcouplings}.
The amplitudes presented here may be useful in 
computing the gluon fusion contamination of the weak boson fusion signal.

The one-loop amplitudes naturally divide into cut-containing $C_n$ and rational parts $R_n$.
We used the double cut unitarity approach of ref.~\cite{Brandhuber:n4}
to derive all the multiplicity results for $C_n$ given in eq.~\eqref{eq:1lhmmCC}.
The rational terms have several sources - first the cut-completion term $CR_n$ which
eliminates the unphysical
poles present in $C_n$, second the direct on-shell recursion  contribution $R_n^D$ and 
third the overlap term $O_n$.   Explicit formulae for these contributions are given in
eqs.~\eqref{eq:1lhmmCR}, \eqref{eq:1lrec} and \eqref{eq:O23}--\eqref{eq:O2i}.
An explicit solution for the four gluon case is given in section~\ref{sec:four}.
We have checked our results in the limit where two of the gluons are collinear, 
in the limit where the scalar becomes soft and against previously known 
results for up to four gluons.

\acknowledgments 
We are grateful to Carola Berger, Zvi Bern, Vittorio Del Duca, Lance Dixon, 
Thomas Gehrmann, David Kosower and Giulia
Zanderighi for stimulating
discussions. The work of SB was supported by Agence Nationale de Recherche grant
ANR-05-BLAN-0073-01. KR thanks David Dunbar for providing highly useful 
Mathematica code, and SLAC for its kind hospitality while some of this work was performed.

\appendix

\section{Spinor conventions}
\label{app:spinorcon}

In the spinor helicity formalism
\cite{Berends:spinhel,deCausmaeker:spinhel,Kleiss:spinhelmass,Gunion:spinhel,Xu:spinhel,Mangano:multipart} an on-shell momentum
of a massless particle, $k_\mu k^\mu=0$, is represented as
\be
k_{\alpha \dot\alpha} \equiv \ k_\mu \sigma^\mu_{\alpha \dot\alpha}
=\ \lambda_\alpha\tilde\lambda_{\dot\alpha} \ ,
\ee
where $\lambda_\alpha$ and $\tilde\lambda_{\dot\alpha}$
are two commuting spinors of positive and negative chirality.
Spinor inner products are defined
by
\be
\langle \lambda,\lambda'\rangle 
= \ \epsilon_{\alpha\beta}\lambda^\alpha\lambda'{}^\beta
 \,, \qquad
[\tilde\lambda,\tilde\lambda'] 
=\ -\epsilon_{\dot\alpha \dot\beta}
\tilde\lambda^{\dot\alpha}\tilde\lambda'{}^{\dot\beta} \,,
\label{Atwo}
\ee
and a scalar product of two null vectors,
$k_{\alpha \dot\alpha}=\lambda_\alpha \tilde\lambda_{\dot\alpha}$ and
$p_{\alpha \dot\alpha}=\lambda'_\alpha\tilde\lambda'_{\dot\alpha}$, becomes
\be \label{scprod}
k_\mu p^\mu =\ - \hf
\langle\lambda,\lambda'\rangle[\tilde\lambda,\tilde\lambda'] \,.
\ee
We use the shorthand $\spa{i}.{j}$ and $\spb{i}.{j}$ for the inner products of
the spinors corresponding to momenta $p_i$ and $p_j$,
\be
\spa{i}.{j} = \langle \lambda_i, \lambda_j \rangle \,,
\qquad
\spb{i}.{j} = [ \tilde\lambda_i, \tilde\lambda_j ].
\ee

For gluon polarization vectors we use
\be
\pol_\mu^\pm(k,\xi) = \pm \frac{ \langle \xi^\mp | \gamma_\mu | k^\mp \rangle
                 }{ \sqrt{2} \langle \xi^\mp | k^\pm \rangle } \,,
\label{HelPol}
\ee
where $k$ is the gluon momentum and $\xi$ is the reference momentum, an
arbitrary null vector which can be represented as the product of two
reference spinors, 
$\xi_{\alpha\dot\alpha}=\xi_{\alpha}\tilde\xi_{\dot\alpha}$.

\section{Scalar integrals}
\label{app:scalarintegrals}

The one-loop functions that appear in the all-orders cut-constructible contribution
${C}_n$ given in eq.~\eqref{eq:1lhmmCC} are defined by,
\begin{align}
	F^{0m}_4(s,t) = \frac{2}{\e^2}
	\bigg[
	&\left (\frac{\mu^2}{-s}\right )^{\e}\FF{1,-\e;1-\e;-\frac{u}{t}} \nonumber\\
	+&\left (\frac{\mu^2}{-t}\right )^{\e}\FF{1,-\e;1-\e;-\frac{u}{s}} \bigg]
	\label{eq:f0m},\\
	F^{1m}_4(P^2;s,t) = \frac{2}{\e^2}
	\bigg[
	&\left (\frac{\mu^2}{-s}\right )^{\e}\FF{1,-\e;1-\e;-\frac{u}{t}} \nonumber\\
	+&\left (\frac{\mu^2}{-t}\right )^{\e}\FF{1,-\e;1-\e;-\frac{u}{s}} \nonumber\\
	-&\left (\frac{\mu^2}{-P^2}\right )^{\e}\FF{1,-\e;1-\e;-\frac{uP^2}{st}} 
	\bigg],
	\label{eq:f1m}\\
	F^{2me}_4(P^2,Q^2;s,t)  = \frac{2}{\e^2}
	\bigg[
	&\left (\frac{\mu^2}{-s}\right )^{\e}\FF{1,-\e;1-\e;
	\frac{
	us
	}{
	P^2Q^2-st
	}} \nonumber\\
	+&\left (\frac{\mu^2}{-t}\right )^{\e}\FF{1,-\e;1-\e;
	\frac{
	ut
	}{
	P^2Q^2-st
	}} \nonumber\\
	-&\left (\frac{\mu^2}{-P^2}\right )^{\e}\FF{1,-\e;1-\e;
	\frac{
	uP^2
	}{
	P^2Q^2-st
	}} \nonumber\\ 
	-&\left (\frac{\mu^2}{-Q^2}\right )^{\e}\FF{1,-\e;1-\e;
	\frac{
	uQ^2
	}{
	P^2Q^2-st
	}}
	\bigg],
	\label{eq:f2me}\\
	F^{1m}_3(s) = \frac{1}{\e^2}&\left (\frac{\mu^2}{-s}\right )^{\e},
	\label{eq:tri}\\
	{\rm Bub}(s) = \frac{1}{\e(1-2\e)}&\left (\frac{\mu^2}{-s}\right )^{\e}.
\end{align}

\bibliographystyle{JHEP-2}
\bibliography{phimm}

\end{document}